\documentclass{article}

\PassOptionsToPackage{numbers,sort&compress}{natbib} %NEW
\usepackage[final]{neurips_2025}

\usepackage[utf8]{inputenc} % allow utf-8 input
\usepackage[T1]{fontenc}    % use 8-bit T1 fonts
\usepackage{hyperref}       % hyperlinks
\usepackage{url}            % simple URL typesetting
\usepackage{booktabs}       % professional-quality tables
\usepackage{amsfonts}       % blackboard math symbols
\usepackage{nicefrac}       % compact symbols for 1/2, etc.
\usepackage{microtype}      % microtypography
\usepackage{xcolor}         % colors
\usepackage{placeins}
\usepackage{amsmath}
\usepackage{amssymb}
\usepackage{mathtools}
\usepackage{hyperref}
\usepackage{standalone}
\usepackage{amsthm}
\usepackage{chemformula}
\usepackage{makecell}
\usepackage{microtype}
\usepackage{graphicx}
\usepackage{booktabs} % for professional tables
\usepackage{subcaption}
\usepackage{tcolorbox}
\usepackage{mathtools}
\usepackage{float}
\usepackage{todonotes}
\theoremstyle{definition}

\usepackage{accents}

\usepackage{tikz}
\usetikzlibrary{arrows.meta, positioning, shapes.geometric, shapes.misc, fit, calc}
\usepackage{amsmath, amssymb} % if not already loaded

\title{Reinforcement Learning for Chemical Ordering in Alloy Nanoparticles}

\author{%
  Jonas Elsborg\textsuperscript{1,2}\;
  Emma L. Hovmand\textsuperscript{1}\;
  \textbf{Arghya Bhowmik}\textsuperscript{1,2}\thanks{Corresponding author: \texttt{arbh@dtu.dk}}\\[1ex]
  \textsuperscript{1}Department of Energy Conversion and Storage, \\ Technical University of Denmark, Kongens Lyngby 2800, Denmark \\
  \textsuperscript{2}CAPeX Pioneer Center for Accelerating P2X Materials Discovery, \\ Technical University of Denmark, Kongens Lyngby 2800, Denmark \\
}

\begin{document}
\maketitle
\begin{abstract}
We approach the search for optimal element ordering in bimetallic alloy nanoparticles (NPs) as a reinforcement learning (RL) problem and have built an RL agent that learns to perform such global optimization using the geometric graph representation of the NPs. To demonstrate the effectiveness, we train an RL agent to perform composition-conserving atomic swap actions on the icosahedral nanoparticle structure. Trained once on randomized \ch{Ag_{X}Au_{309-X}} compositions and orderings, the agent discovers previously established ground state structure. We show that this optimization is robust to differently ordered initialisations of the same NP compositions. We also demonstrate that a trained policy can extrapolate effectively to NPs of unseen size. However, the efficacy is limited when multiple alloying elements are involved. Our results demonstrate that RL with pre-trained equivariant graph encodings can navigate combinatorial ordering spaces at the nanoparticle scale, and offer a transferable optimization strategy with the potential to generalize across composition and reduce repeated individual search cost.
\end{abstract}

% NEW
\vspace{0.5em}

\section{Introduction}
Metallic nanoparticles (NPs) are widely used as heterogeneous (electro)catalysts because their high site-to-volume ratios and tunable active sites enable exceptional reactivity~\citep{narayan2019metal, mcintyre2025computational}. The elemental composition, size, shape, and atomic ordering of NPs directly influence their catalytic properties and stability by determining the distribution and nature of active sites on the surface~\citep{loevlie2023demystifying}. Determining the atomic structure of NPs is therefore a central objective in designing efficient and selective catalysts, and several computational strategies have been developed to resolve their ground-state atomic arrangements~\citep{loevlie2023demystifying}. However, NP structure search faces two central challenges: the high cost of evaluating the total energy of a given atomic structure/ordering to assess its stability, and the limited efficiency of algorithms for exploring the structure/ordering space using this energy evaluator. As a result, most NP structure searches rely on inexpensive classical potentials~\citep{lysgaard2014genetic, zhang2015global, cerbelaud2011optimization}, while first-principles Density Functional Theory (DFT) is generally too computationally expensive to use in the search itself and is instead reserved for validating a limited set of candidate structures~\citep{zhang2015global,negreiros2010structures}. Recently, machine learning (ML) potentials have been used to accelerate the energy evaluation step while retaining near-DFT accuracy in energetics~\citep{han2022unfolding, alvarez2021neural, ojha2024ann}. 
The search space for possible atomic configurations grows combinatorially with cluster size, and constructing effective search algorithms is equally crucial and can be paired with all different energy evaluation methods.
Mixed-integer programming (MIP)~\citep{larsen2018rich}, Monte Carlo (MC) sampling~\citep{artrith2015grand, pirart2019reversed, bochicchio2014chemical}, and basin hopping (BH)~\citep{gould2014segregation, cheng2013structure, rondina2013revised, bochicchio2013morphological} techniques have been used to determine provably optimal structures (within the accuracy of the energy/fitness function). Genetic algorithms (GAs) explore the vast combinatorial configuration space by iteratively evolving a population of candidate structures, and are generally more efficient and popular~\citep{chen2008energetic, froemming2009optimizing, loevlie2023demystifying, lysgaard2014genetic, lysgaard2015dft, han2022unfolding, rodrigues2008global, zhang2015global, tao2017stable, dean2020rapid, liu2016structural, rapallo2005global}. For each GA run, one must evaluate the fitness (energy) of many candidates over numerous generations. However, GA and other classical methods like MC are intractable for larger systems, and a significant drawback from these methods is the non-transferability and lack of generalization across different compositions, forcing independent optimization runs for each elemental composition or NP size separately. 

\textcolor{black}{Starting from seminal work by Hammer et al. using image based reinforcement learning (RL) ~\citep{jorgensen2019atomistic, mortensen2020atomistic, meldgaard2020structure, christiansen2020gaussian, ager2022generating},} recent work has turned to RL as a unified optimization approach for atomic structures~\citep{meldgaard2020structure, ahmed2025materials}. An RL agent can be trained to construct molecules or nanoparticle structures atom-by-atom~\citep{christiansen2020gaussian, ager2022generating, simm2020reinforcement,elsborg2023equivariant}, or even fragment-by-fragment~\citep{flam2022scalable}. Another approach is to train RL agents to modify an existing structure by treating the configuration space as a labelled graph, which the RL agent learns to reconfigure~\citep{elsborg2024artisan}. The benefit of a trained RL policy is that it can rapidly propose near-optimal configurations without performing an exhaustive search from scratch for each structure or composition. So far, this approach has been explored only for nanoparticles with small number of atoms~\citep{elsborg2023equivariant,ahmed2025materials}. In this work, we built and deployed a reinforcement learning framework that couples an equivariant encoder~\citep{rhodes2025orbv3atomisticsimulationscale} with a reinforcement learning model trained via proximal policy optimization (PPO) to perform composition-preserving structure manipulation actions on the atomic graph to optimize ordering. Trained once, the agent reproduces a set of provably exact ground state orderings of Mackay-icosahedral \ch{Ag_{X}Au_{309-X}} nanoparticles.

%Problem and Setting. Learning the configuration space of 309-atom Ag–Au Mackay icosahedra via discrete atom–swap actions to minimize potential energy. Key Idea. Couple a frozen/tunable equivariant ORB encoder with a lightweight PPO actor–critic to choose focus/swap atoms, conditioned on time/horizon. Contributions. (i) Swap-centric RL environment with progress-shaped rewards; (ii) ORB features + predicted-vector cues for action scoring; (iii) DDP-synchronized on-policy collection and KL-gated PPO; (iv) curated EXP1 dataset + fixed 8-composition validation set. Scope and Assumptions. Non-periodic nanoparticles; Ag–Au only; energy as sole objective; relaxation per-step or terminal; no kinetic constraints.

\section{Methods}\label{sec:methods}
\subsection{Nanoparticle optimisation as a reinforcement learning problem}
We formulate the search for a minimum-energy nanoparticle structure as a sequential decision-making task, i.e., a Markov Decision Process (MDP). In this MDP, a state $s_{t}$ represents an atomic configuration of a NP (positions and types of all atoms) at time step $t$ in the MDP. In our setting, an action $a_t$ at time $t$ corresponds to a modification of the NP structure resulting from swapping the positions $\mathbf{x_A}=(x_{A}, y_A,z_A)$ and $\mathbf{x_B}=(x_{B}, y_B,z_B)$ of two atoms A and B. The environment computes the potential energy of the resulting structure and returns a reward $r_{t}=E(s_{t})-E(s_{t+1})$ based on the energy change between the energy of the pre-swap state, $E(s_{t})$, and that of the post-swap state, $E(s_{t+1})$. After each action, we also perform a local geometry relaxation. \\
Thus, if $r_{t}>0$, the atomic swap resulted in a lower-energy state at $t+1$ than at $t$. This approach incentivises the agent to move toward more stable configurations. It also fully satisfies the Markov property (the energy outcome depends only on the current atomic arrangement) and provides a dense scalar reward. Maximising the return $G$, i.e. the cumulative reward, over a horizon of $H$ actions corresponds to minimising the NP’s energy. This results in a sum that signals the overall energy difference between the initial state and the final state (assuming no discount factor is used, see Appendix~\ref{app:mdp}). The formulation allows the agent to explore the configurational space in a guided manner since it learns a policy that, at each step, chooses how to alter the NP to reach a lower-energy state at the end. 
By maximising G, the agent can discover non-trivial sequences of modifications that yield large net energy drops, even if some intermediate steps may lead to an increase in energy. This ability of RL to handle long-term credit assignment (assigning credit to actions that lead to benefits much later) is a key advantage in navigating complex energy landscapes with many local minima. 

Recent studies have demonstrated that policy-based RL agents can efficiently navigate such energy landscapes to find stable, low-energy structures of molecules, metal clusters and bulk materials~\citep{simm2020reinforcement, flam2022scalable, elsborg2023equivariant, elsborg2024artisan}. To evaluate the energy of NPs, we use the Effective Medium Theory (EMT) potential - a semi-empirical many-body potential~\citep{jacobsen1996semi}. Even though EMT has relatively cheap energy evaluations, the search for global minima can be very expensive as the number of chemical orderings grows combinatorially. For a fixed shape with $N$ sites and bimetallic composition $A_x B_{N-x}$, the labellings scale as $\binom{N}{x}$. For example, in a 309-atom icosahedral NP (\ch{Ag_{X}Au_{309-X}}) with elemental composition of \ch{Ag162Au147}, we get $\binom{309}{162}\approx 3.3 \times 10^{91}$ chemical orderings, and symmetry reduces this only by a modest constant factor~\citep{dean2020rapid}.
\subsection{RL Algorithm: Actor–Critic with PPO and KL Regularisation}
We adopt an actor–critic version of the proximal policy optimisation algorithm~\citep{schulman2017proximal} (PPO) to train the agent’s policy. The actor is the policy $\pi_\theta(a\mid s)$, parametrised by weights $\theta$, used to predict a distribution over actions $a_{t}$ given state $s_{t}$ at time $t$. The critic is a value function $V_\phi(s_{t})$, with parameters $\phi$, and it estimates the expected cumulative reward (proportional to the expected energy reduction) from $s_{t}$. By teaching the critic to predict the state values, the actor can learn to select actions that yield high returns as judged by the critic’s feedback. We use generalised advantage estimation~\citep{schulman2015high} (GAE) to compute the advantage of an action $a_{t}$ compared to default behaviour (expectation under the value function). All algorithm details are in Appendix \ref{appendix:algorithm}. \\
Due to our definition of MDP as a sequence of atomic swap actions, we define a single atom swap action as the choice of an atom pair $a_t=(i_t,j_t)$. Thus, at each time step $t$, the agent picks an \textit{anchor} atom $i$ which is then paired with a suitable \textit{partner} atom indexed by $j$ (Figure \ref{fig:step_fig}). 
% In your preamble:
\usetikzlibrary{arrows.meta,positioning}
\tikzset{
  anchorAtom/.style={draw=blue!70!black, ultra thick, circle, minimum size=4mm},
  partnerAtom/.style={draw=red!75!black, ultra thick, circle, minimum size=4mm},
  callout/.style={draw, rounded corners, thick, fill=white, opacity=0.9, text opacity=1, inner sep=3pt},
}

% In your document:
\begin{figure}[h!]
  \centering
  \begin{tikzpicture}
    % Place the image
    \node[inner sep=0] (img) {\includegraphics[width=.25\textwidth]{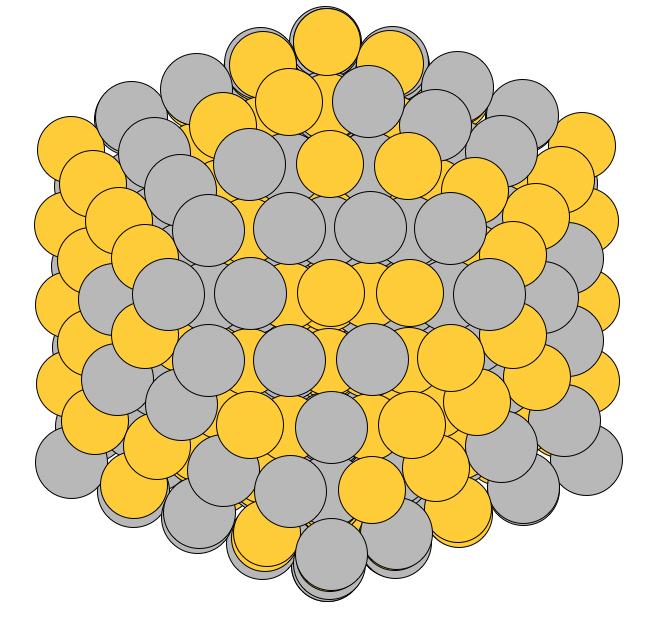}};

    % Use normalized (0..1) coordinates over the image
    \begin{scope}

      % === Choose coordinates (adjust these two pairs) ===
      \def\ax{-0.65}\def\ay{0.45}   % anchor (center-ish)
      \def\px{0.63}\def\py{-0.25}   % partner (lower-right-ish)

      % Anchor and partner markers
      \node[anchorAtom]  (anchor)  at (\ax,\ay) {};
      \node[partnerAtom] (partner) at (\px,\py) {};

      % Rings so the original colors shine through a bit
      \draw[blue!70!black, ultra thick]  (\ax,\ay) circle [radius=0.035];
      \draw[red!75!black,  ultra thick]  (\px,\py) circle [radius=0.035];

      % Labels (auto-positioned; nudge with xshift/yshift if needed)
      \node[callout, above left=3pt of anchor]  {\textbf{anchor} $i_t$};
      \node[callout, below right=3pt of partner] {\textbf{partner} $j_t$};

      % Swap arrows (two-headed effect with two curved arrows)
      \draw[-{Latex[length=3mm,width=2mm]}, very thick, blue!70!black]
        (anchor) to[bend left=20] (partner);
      \draw[-{Latex[length=3mm,width=2mm]}, very thick, red!75!black]
        (partner) to[bend left=20] (anchor);

      % Per-step callout
      \node[callout, align=center, anchor=north]
        at ($(img.south)+(0,-3mm)$)
        {\(\textbf{swap}~(i_t,j_t) \rightarrow \textbf{relax} \rightarrow
          \textbf{reward}~r_t=E(s_{t})-E(s_{t+1})\)};
    \end{scope}
  \end{tikzpicture}

  \caption{Agent–environment step on the Ag/Au NP: highlight the chosen \textbf{anchor} $i$ and \textbf{partner} $j$, swap species, relax, compute reward, and continue. In Figure \ref{fig:full_traj_snaps} in Appendix \ref{app:additional results}, we show snapshots from a full episode of anchor-partner swaps for a trained model. }
  \label{fig:step_fig}
\end{figure}

We use a factorized actor which acts on features distilled from a frozen ORB-v3 atomic graph encoder~\citep{rhodes2025orbv3atomisticsimulationscale}. We use two neural network policy heads:
\begin{equation}
    \pi_\theta(a_t\mid s_t)
=\pi_{\theta_a}(i_t\mid s_t)\;\pi_{\theta_p}(j_t\mid s_t,i_t),\quad \theta=(\theta_a,\theta_p),
\end{equation}
where $\theta_a$ are the features of the \textit{anchor} policy neural network, and $\theta_p$ are the features of the \textit{partner} policy neural network, which is conditioned on the chosen anchor atom indexed by $i_t$ (See Appendix \ref{app:policy_factor_details} for details on how these policy networks are constructed).

\section{Experimental Results}
Icosahedral NPs are used here to demonstrate the effectiveness of the proposed learning framework. Catalytic alloy NPs are often icosahedral, which are interesting due to the presence of a high density of edges/vertices, creating strained, low-coordination sites that can strongly modulate reaction intermediate adsorption energies. The discrete “magic-number” shells (N=13, 55, 147, 309, 561,...) also promote composition-dependent surface segregation and core–shell ordering, phenomena directly linked to catalytic selectivity and activity, and thus are critical to resolve accurately~\citep{ferrando2008nanoalloys}. As a testbed for our model, icosahedra are ideal: they have clear, well-studied ground-state motifs, yet they have an enormous configuration space with many energetically near-degenerate arrangements~\citep{molayem2011global, molayem2011theoretical, zhang2006structure}. This creates a challenging test case, and effective policies must both discover the right shell assignments and avoid combinatorial traps. Their standardized sizes let us probe generalization across composition with controlled geometry and later test size generalization.

\tikzset{
  box/.style={draw, rounded corners, thick, align=center, inner sep=6pt, fill=gray!5},
  proc/.style={box, fill=blue!5},
  data/.style={box, fill=green!8},
  note/.style={box, fill=yellow!12},
  loss/.style={box, fill=red!6},
  line/.style={-Latex, thick},
  tinytext/.style={font=\scriptsize}
}

\begin{figure}[ht!]
    \centering
    \begin{tikzpicture}[node distance=10mm and 10mm, font=\small]
        
% === Main vertical pipeline (top -> bottom) ===
\node[data, text width=62mm, minimum height=14mm] (NP) {%
  \textbf{Nanoparticle}\\
  Ag/Au 309-atom icosahedron\\
  $\{ \mathbf{x}_i, Z_i\}_{i=1}^{309},~E_t$
};

\node[proc, below=10mm of NP, text width=62mm, minimum height=14mm] (Graph) {%
  \textbf{ASE $\to$ Atom Graph}\\
  Senders, Receivers, Edges
};

\node[proc, below=10mm of Graph, text width=55mm, minimum height=18mm] (Enc) {%
  \textbf{Encoder (ORB-v3)}\\
  Node features $\mathbf{h}_i$, Edge features $\mathbf{e}_i$\\
  Force vectors $\mathbf{v}_i$
};

\node[proc, below=12mm of Enc, text width=55mm, minimum height=23mm] (AC) {%
  \textbf{Actor--Critic}\\[0.6mm]
  \textbf{Anchor head} $\pi_a(i| s_t)$\quad \\
  \textbf{Partner head} $\pi_p(j|i,s_t)$ (masked)\\
  \textbf{Critic} $V_\phi(s_t)$
};

\node[proc, below=12mm of AC, text width=55mm, minimum height=18mm] (Env) {%
  \textbf{Environment step}\\
  Swap species at $(i,j)$; \\
  Quick relax (L-BFGS+EMT)\\
  Update $s_{t}\!\to\!s_{t+1}$
};

\node[data, below=10mm of Env, text width=60mm, minimum height=16mm] (Buf) {%
  \textbf{On-policy buffer}\\
  Store state snapshot $\tau_t$
};

\node[loss, below=12mm of Buf, text width=60mm, minimum height=15mm] (PPO) {%
  \textbf{GAE + PPO update}\\[0.6mm]
  Compute $\hat{A}$ and $\mathcal{L}^{\mathrm{CLIP}}(\theta)$\\
  KL early stop if $\text{KL} > 1.5\,\tau$
};

% --- flow arrows ---
\draw[line] (NP) -- (Graph);
\draw[line] (Graph) -- (Enc);
\draw[line] (Enc) -- (AC);
\draw[line] (AC) -- node[right, tinytext]{Sample (training) or $\epsilon$-greedy argmax (final evaluation) $(i,j)$} (Env);
\draw[line] (Env) -- node[right, tinytext]{Store transition} (Buf);
\draw[line] (Buf) -- (PPO);

% === Side notes: features and reward ===
\node[note, right=18mm of Enc, text width=58mm] (AnchorFeat) {%
  \textbf{Anchor features} for atom $i$:\\[0.4mm]
  $\boldsymbol{\phi}^{(a)}_i(t)=\big[\ \mathbf{h}_i,\ \boldsymbol{\chi}_i,\ \psi(t,H)\ \big]$
};
\draw[line] (Enc.east) -- (AnchorFeat.west);

\node[note, right=18mm of AC, text width=68mm] (PartnerFeat) {%
  \textbf{Partner features} for candidate $j$ given Anchor $i$:\\[0.6mm]
  $\boldsymbol{\phi}^{(p)}_{j\mid i}(t)
=
\big[\ \mathbf{h}_j,\ \mathbf{h}_i,\ \bar{\mathbf{e}}_{i\to j},\ d_{ij},\ a_{ij},\ \psi(t,H)\ \big]$\\[0.6mm]
  mask: $Z_j=Z_i \ \Rightarrow\ \text{logits}=-10^9$
};
\draw[line] (AC.east) -- (PartnerFeat.west);

\node[note, right=18mm of Env, text width=60mm] (Reward) {%
  \textbf{Per-step reward (with relaxation)}:\\[0.3mm]
  $r_t = E(s_{t}) - E(s_{t+1})$,\quad $N{=}309$\\
  episode horizon $T = N$
};
\draw[line] (Env.east) -- (Reward.west);

\node[data, below=8mm of PPO, text width=86mm, align=left] (Comps) {%
\textbf{Validation compositions (Ag/Au, $N{=}309$)}:\\
$8\times {Ag_{X}Au_{309-X}}, \ \ \ \ X=\{296,228,205,162,126,75,25,13\}$
};
    \end{tikzpicture}
    \caption{PPO implementation and model flow for the proposed composition generalized model for global nanoparticle atomic ordering optimization.}
    \label{fig:agau309-ppo}
\end{figure}
\begin{figure}[ht]
    \centering
    % Row 1
    \begin{subfigure}{0.4\textwidth} % top-left cell
        \centering
        \raisebox{0.25cm}{\includegraphics[width=0.35\linewidth]{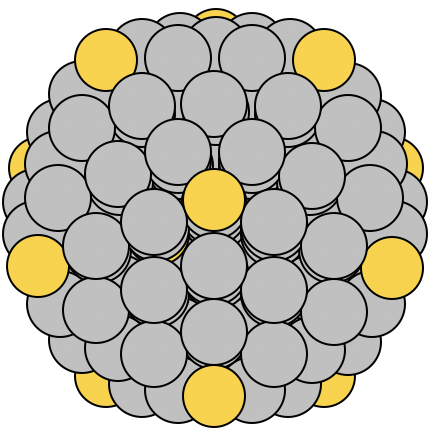}}
        \includegraphics[width=0.45\linewidth]{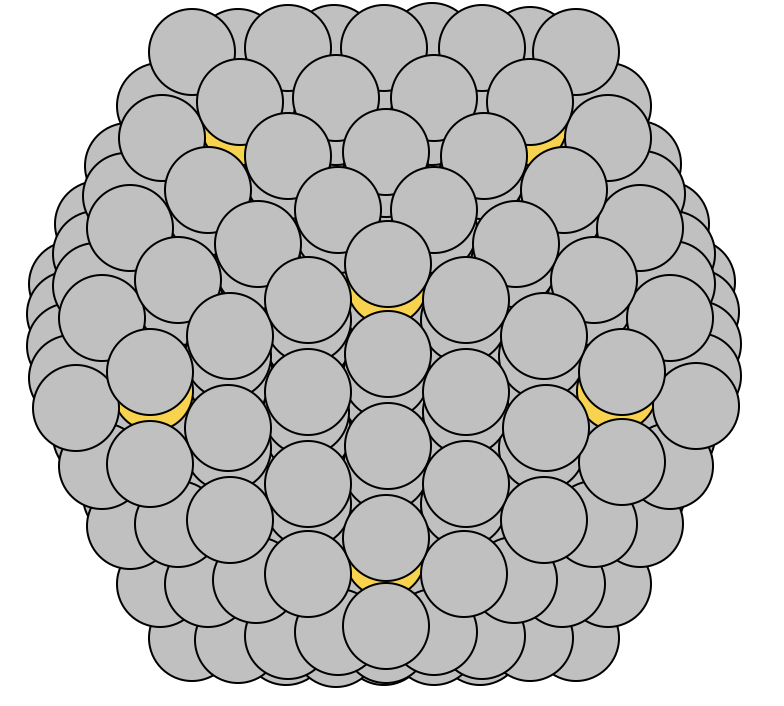}
        \caption{\ch{Ag296Au13}}
        \label{subfig:ih1}
    \end{subfigure}
    \hfill
    \begin{subfigure}{0.4\textwidth} % top-right cell
        \centering
        \raisebox{0.35cm}{\includegraphics[width=0.35\linewidth]{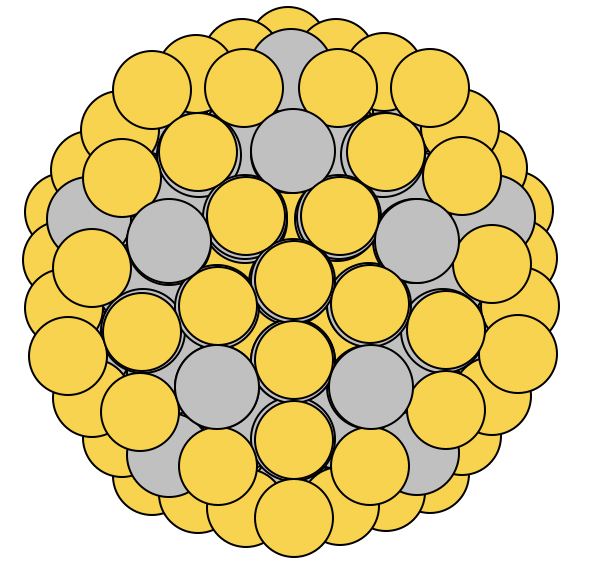}}
        \includegraphics[width=0.43\linewidth]{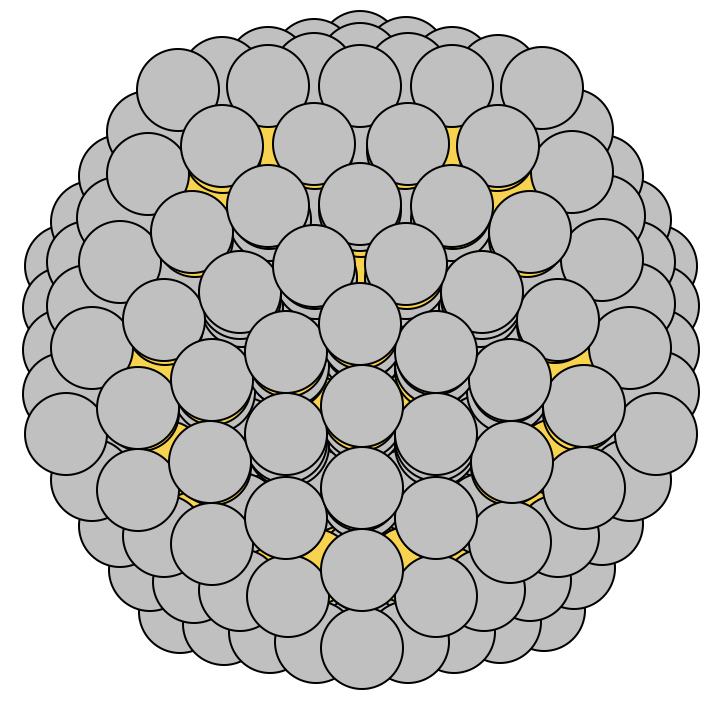}
        \caption{\ch{Ag228Au81}}
        \label{subfig:ih2}
    \end{subfigure}    
    % Row 2
    \begin{subfigure}{0.4\textwidth} % bottom-left cell
        \centering
        \raisebox{0.4cm}{\includegraphics[width=0.25\linewidth]{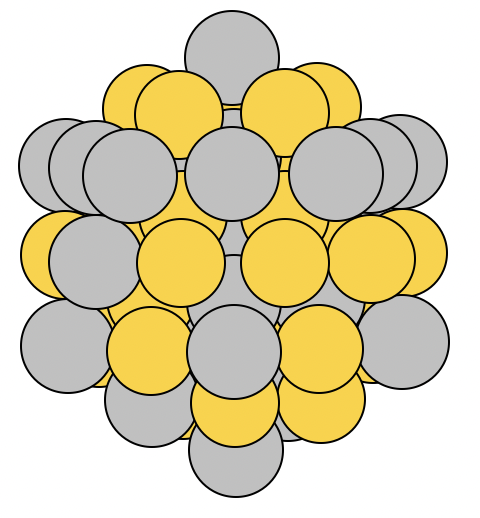}}
        \includegraphics[width=0.43\linewidth]{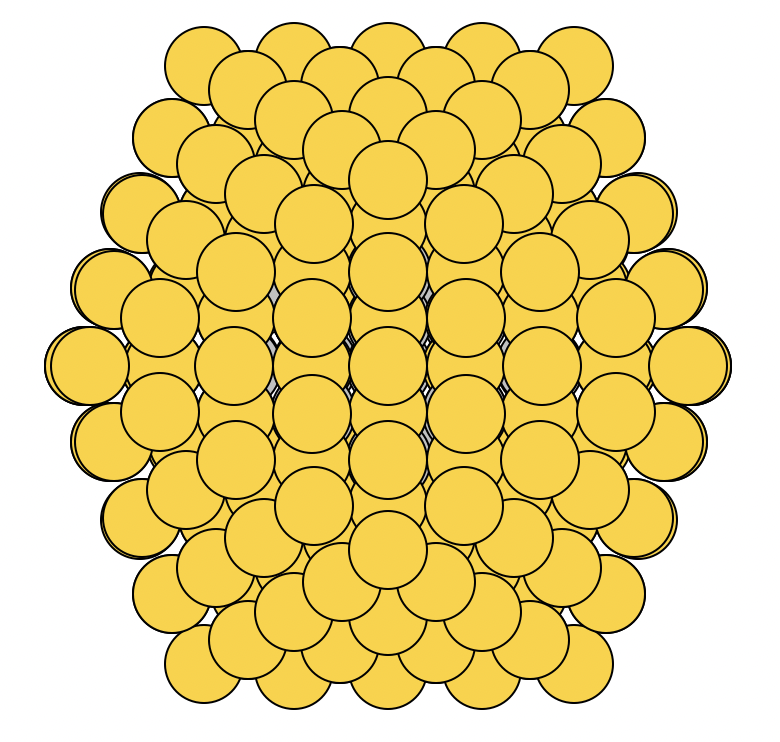}
        \caption{\ch{Ag25Au284}}
        \label{subfig:ih7}
    \end{subfigure}
    \begin{subfigure}{0.4\textwidth} % bottom-right cell
        \centering
        \raisebox{0.4cm}{\includegraphics[width=0.25\linewidth]{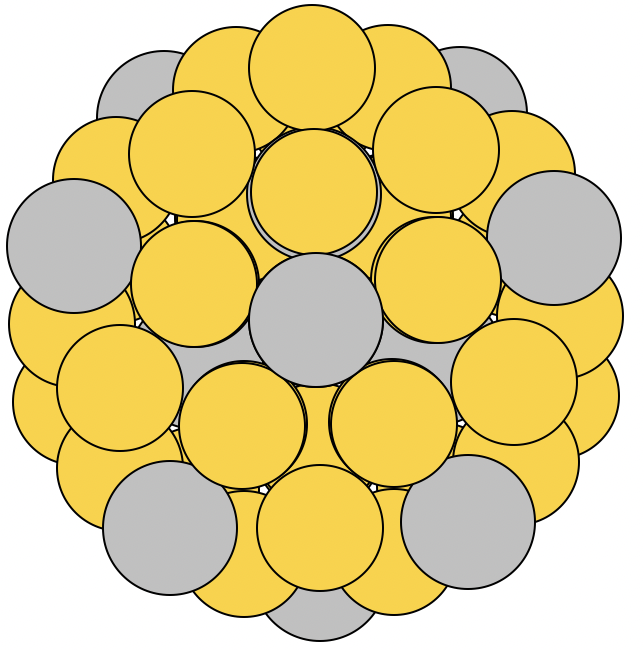}}
        \includegraphics[width=0.43\linewidth]{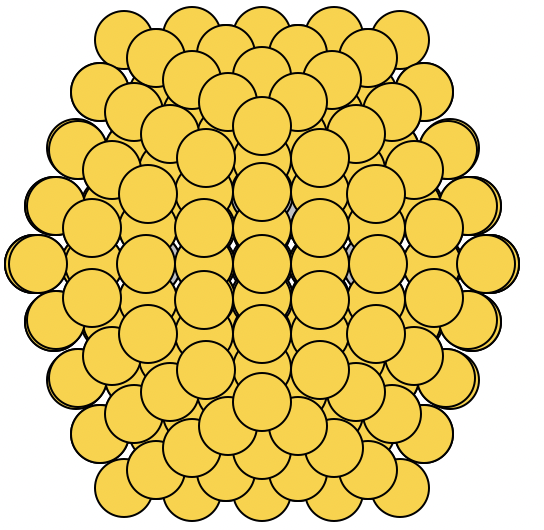}
        \caption{\ch{Ag13Au296}}
        \label{subfig:ih8}
    \end{subfigure}
    \begin{subfigure}{0.2\textwidth} % bottom-left cell
        \centering
        \includegraphics[width=0.8\linewidth]{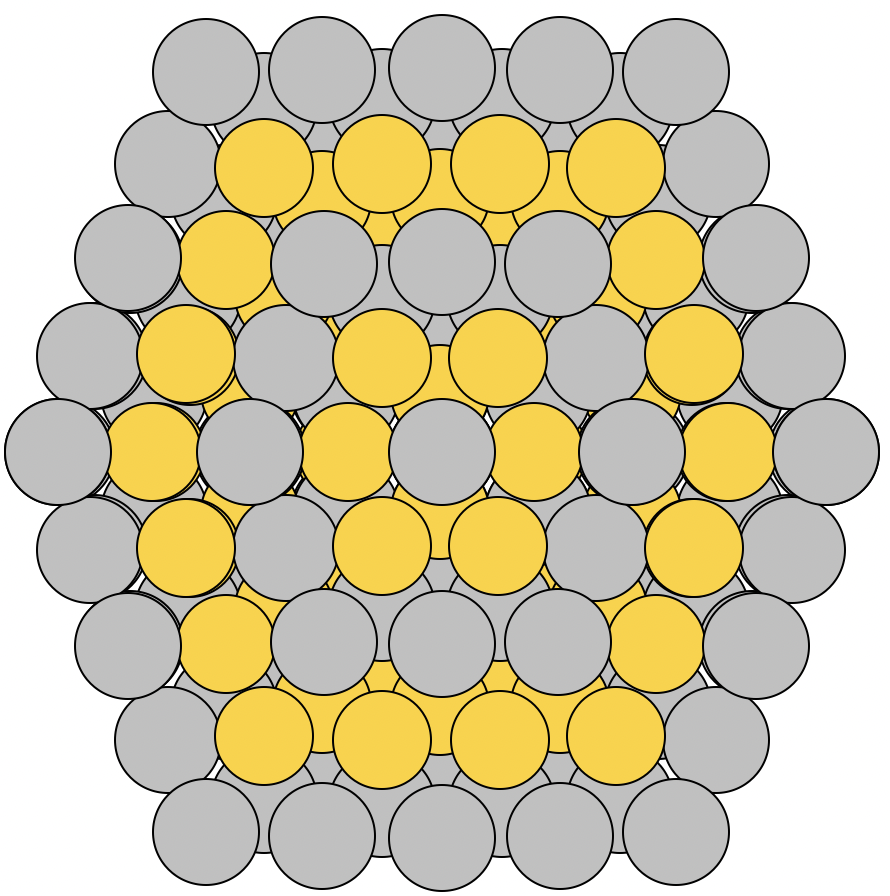}
        \caption{\ch{Ag205Au104}}
        \label{subfig:ih3}
    \end{subfigure}
    \begin{subfigure}{0.2\textwidth} % bottom-left cell
        \centering
        \includegraphics[width=0.8\linewidth]{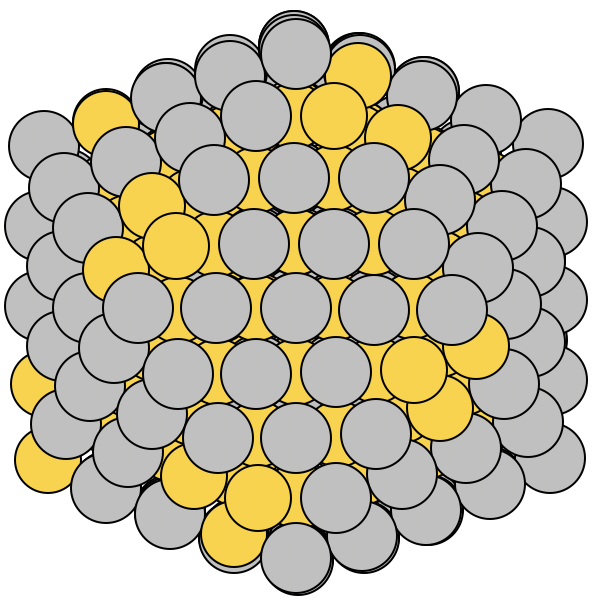}
        \caption{\ch{Ag162Au147}}
        \label{subfig:ih4}
    \end{subfigure}
    \begin{subfigure}{0.2\textwidth} % bottom-right cell
        \centering
        \includegraphics[width=0.8\linewidth]{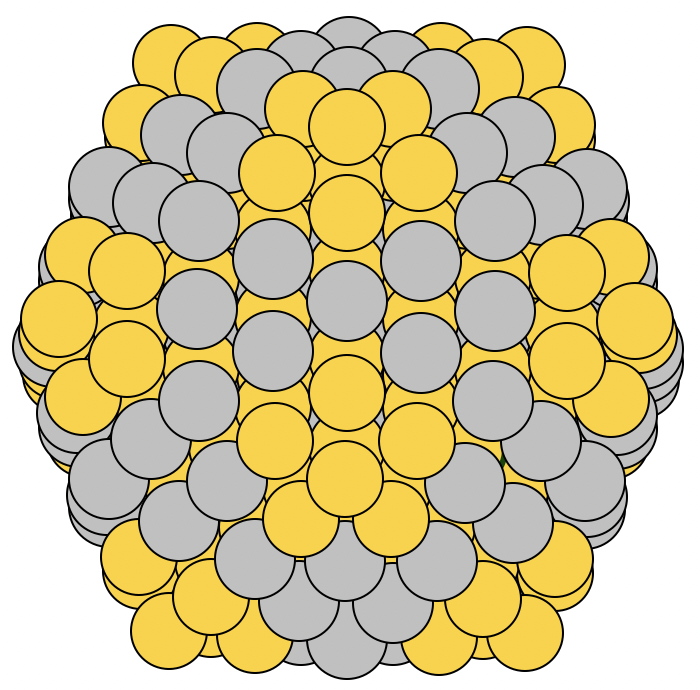}
        \caption{\ch{Ag126Au183}}
        \label{subfig:ih5}
    \end{subfigure}
    \begin{subfigure}{0.2\textwidth} % bottom-left cell
        \centering
        \includegraphics[width=0.8\linewidth]{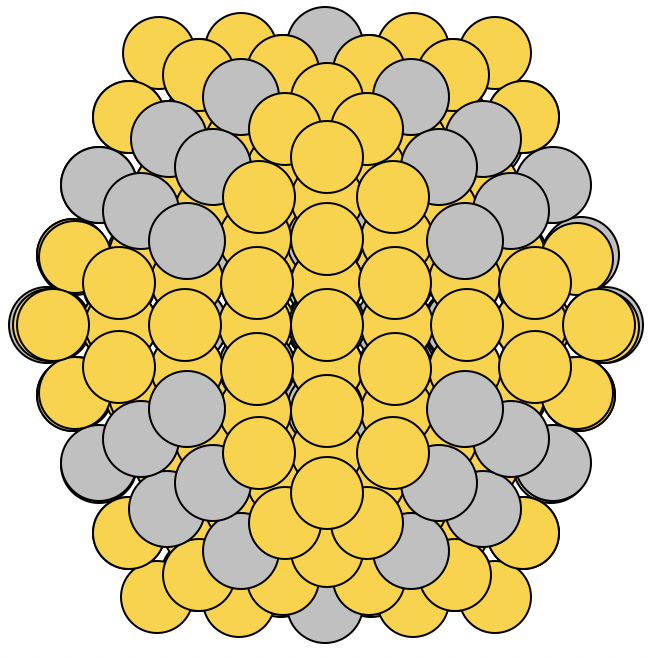}
        \caption{\ch{Ag75Au234}}
        \label{subfig:ih6}
    \end{subfigure}
    \caption{Lowest energy structures found by the trained agent for (a)-(h) the eight test compositions. For the first four (a)-(d) both the internal core structure (left) and outside shell structure (right) are provided.}
    \label{fig:results}
\end{figure}

\newpage
\paragraph{Testing}For final testing of the generalized agent, we treat the provably lowest energy structures with respect to the EMT potential established with MIP solutions of \ch{Ag_{X}Au_{309-X}} icosahedral nanoparticles of \citet{larsen2018rich} as reference targets, and use eight compositions whose ground state orderings were specifically reported there. At the end of training, each GPU samples a randomized configuration of these formulas, and the agent is tasked with solving these structures to find the minimum energy using an evaluation horizon $H_{eval}=2\times N_{atoms}=618$. \\
In Figure \ref{fig:results}, we show the single ordering (out of 8 - one for each GPU) found for each composition, which had the lowest final energy.

\subsection{Optimiser policy can generalise across composition (Experiment-1)}
Current state-of-the-art models require running the ordering search individually from scratch for each elemental composition for an NP motif. A significant breakthrough \textcolor{black}{
could be achieved by learning a policy that generalizes across elemental compositions.} 

\paragraph{Training} To evaluate composition generalisation of the model, we trained for 100 epochs $\times \ 300$ episodes on 309-atom icosahedra with fully randomised \ch{Ag_{X}Au_{309-X}} compositions and orderings (no ordering repeats). Training uses eight 24 GB RTX 3090 GPUs, each sampling distinct nanoparticles (varied by ordering and elemental composition) per episode. The horizon is $H_{\text {train }}=309$, and for each step, the agent selects an anchor-partner pair, swaps them, relaxes the structure, and receives a reward $r_t=E(s_t)-E(s_{t+1})$. Because all states are similarly relaxed, $r_t$ reflects only ordering improvements. We optimise actor-critic networks with PPO + GAE while keeping the ORB-v3 encoder frozen (Figure \ref{fig:agau309-ppo}, see Appendix \ref{appendix:algorithm} for further details). We show training curves in Figure \ref{fig:training_curves} in Appendix \ref{app:training_curves}.

\paragraph{Results} For the structures dominated by a single species in Figures \ref{subfig:ih1}-\ref{subfig:ih8}, we show both an interior view in the left figures and an exterior view of the shell in the right figures. For structures dominated by \ch{Ag} (Figures \ref{subfig:ih1} and \ref{subfig:ih2}), the agent prefers an Ag-shell with Au atoms placed either at the subsurface corners (Figure \ref{subfig:ih1}) or at the subsurface edges (Figure \ref{subfig:ih2}). These structures are the same as the ground state structures found by \citet{larsen2018rich}. For structures dominated by \ch{Au} (Figures \ref{subfig:ih7} and \ref{subfig:ih8}), the agent prefers an Au-shell with Ag atoms placed in the innermost 3 shells, with no Ag atoms in shells 4 and 5. The patterns of these structures are also similar to those found in \citet{larsen2018rich}, except for a few atoms. However, the agent correctly places an \ch{Ag} atom in the centre of these nanoparticles (in fact, all nanoparticles have a correctly placed central atom). For the more balanced compositions of Figures~\ref{subfig:ih3}-\ref{subfig:ih6}, the agent finds the same structures as those in~\citet{larsen2018rich}, including the perfectly ordered "onion-shell" structure for \ch{Ag205Au104}. The agent also finds the correct "flower-like" surface decoration of \ch{Ag126Au183} (Figure~\ref{subfig:ih5}), as well as the \ch{Ag162Au147} surface decoration with 3-atom clusters of \ch{Au} (Figure \ref{subfig:ih4}), and the \ch{Ag75Au234} surface decoration with 6-atom clusters of \ch{Ag} (Figure~\ref{subfig:ih6}). These results affirm the suitability of our model towards generalization across the chemical composition of a particular-sized icosahedral NP, lowering the cost of atomic ordering search through amortization. 

\begin{figure*}[ht!]
\centering
% ---------- Row 1 ----------
\begin{subfigure}[t]{0.49\textwidth}
  \centering
  \includegraphics[width=0.44\linewidth]{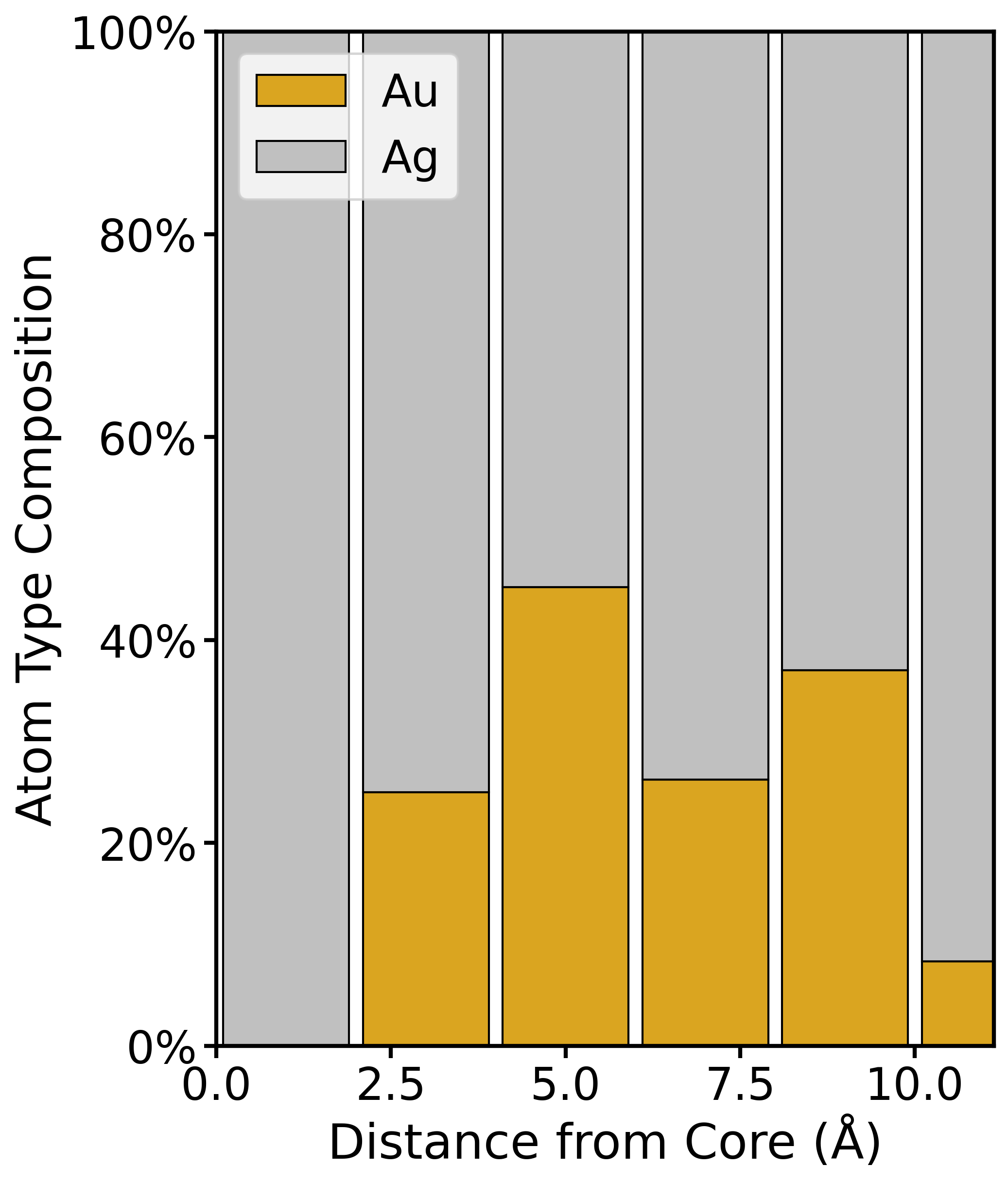}\hfill
  \includegraphics[width=0.44\linewidth]{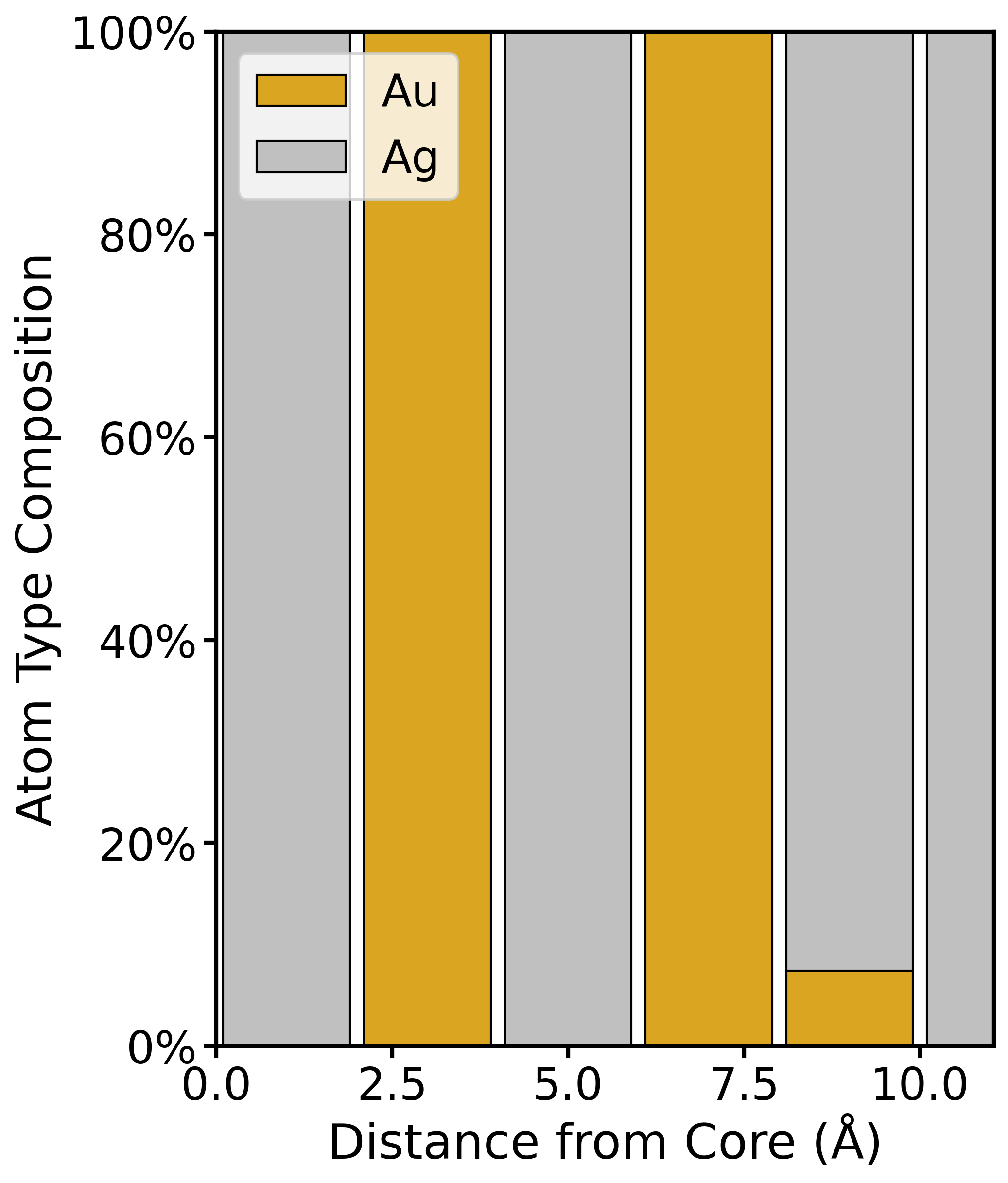}
  \subcaption{Best ($52.892377 \ \mathrm{eV} \rightarrow  49.276259 \ \mathrm{eV}$)}
\end{subfigure}\hfill
\begin{subfigure}[t]{0.49\textwidth}
  \centering
  \includegraphics[width=0.44\linewidth]{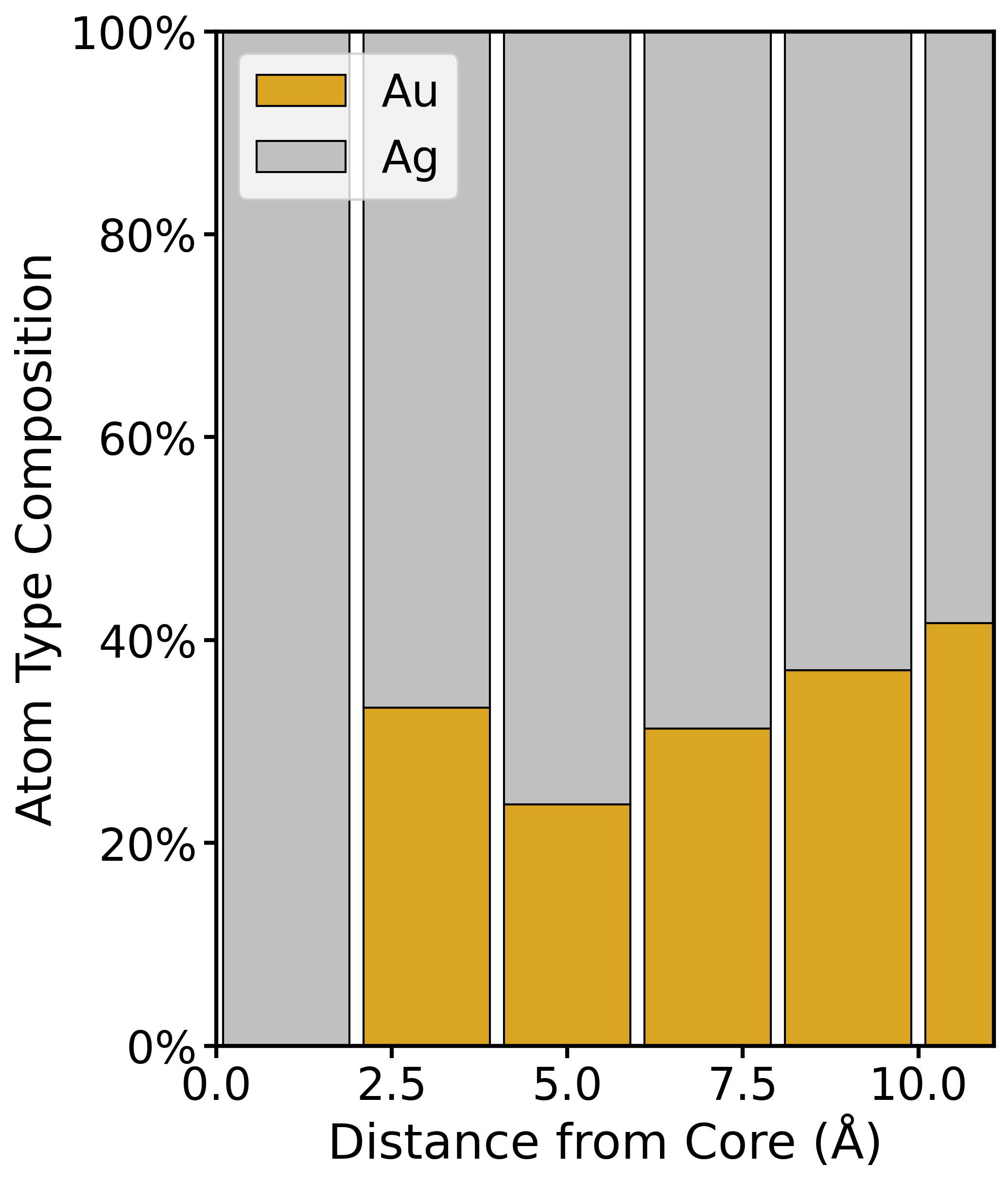}\hfill
  \includegraphics[width=0.44\linewidth]{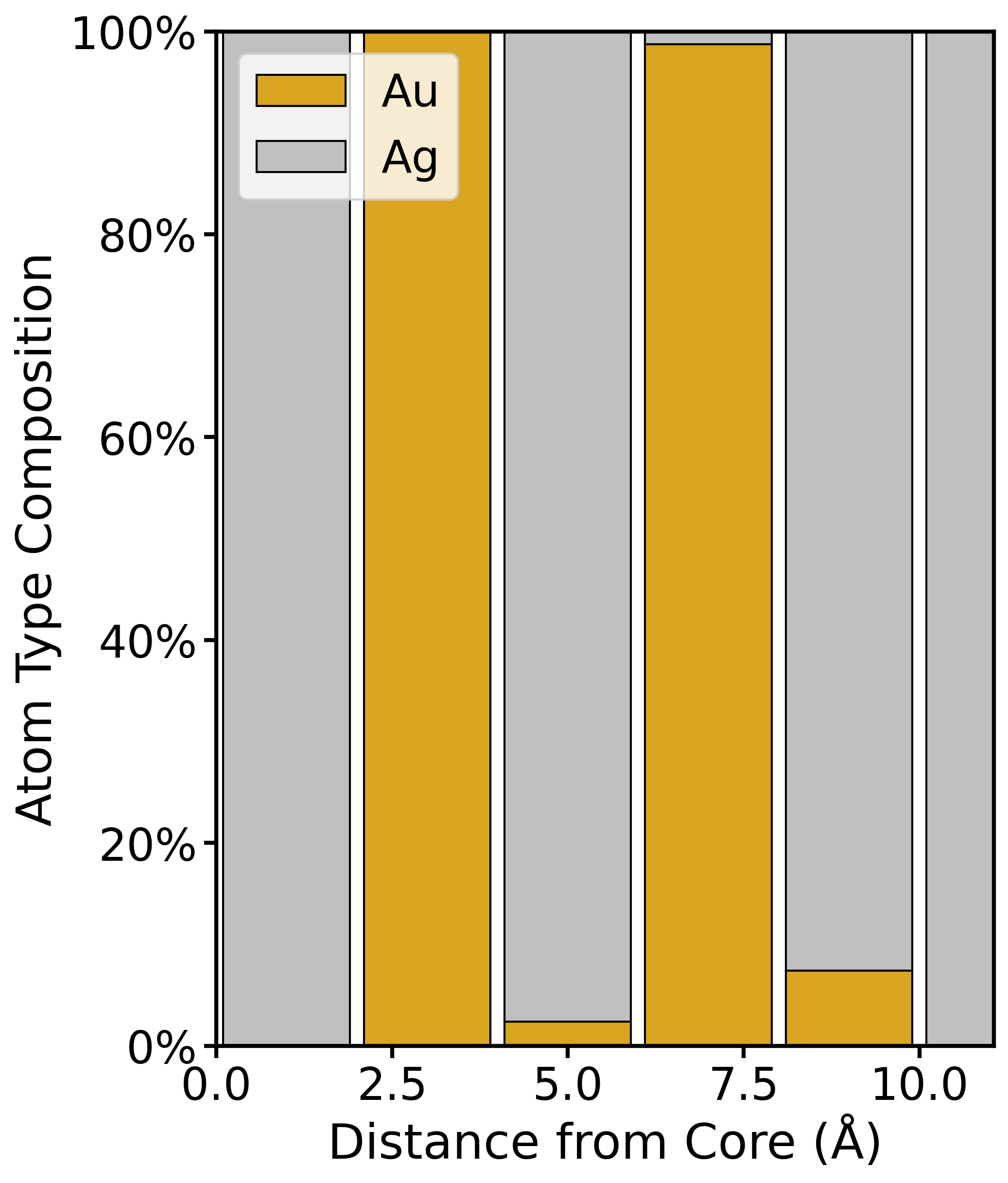}
  \subcaption{2nd best ($52.432193 \ \mathrm{eV} \rightarrow  49.276264 \ \mathrm{eV}$)}
\end{subfigure}
% ---------- Row 2 ----------
\begin{subfigure}[t]{0.49\textwidth}
  \centering
  \includegraphics[width=0.44\linewidth]{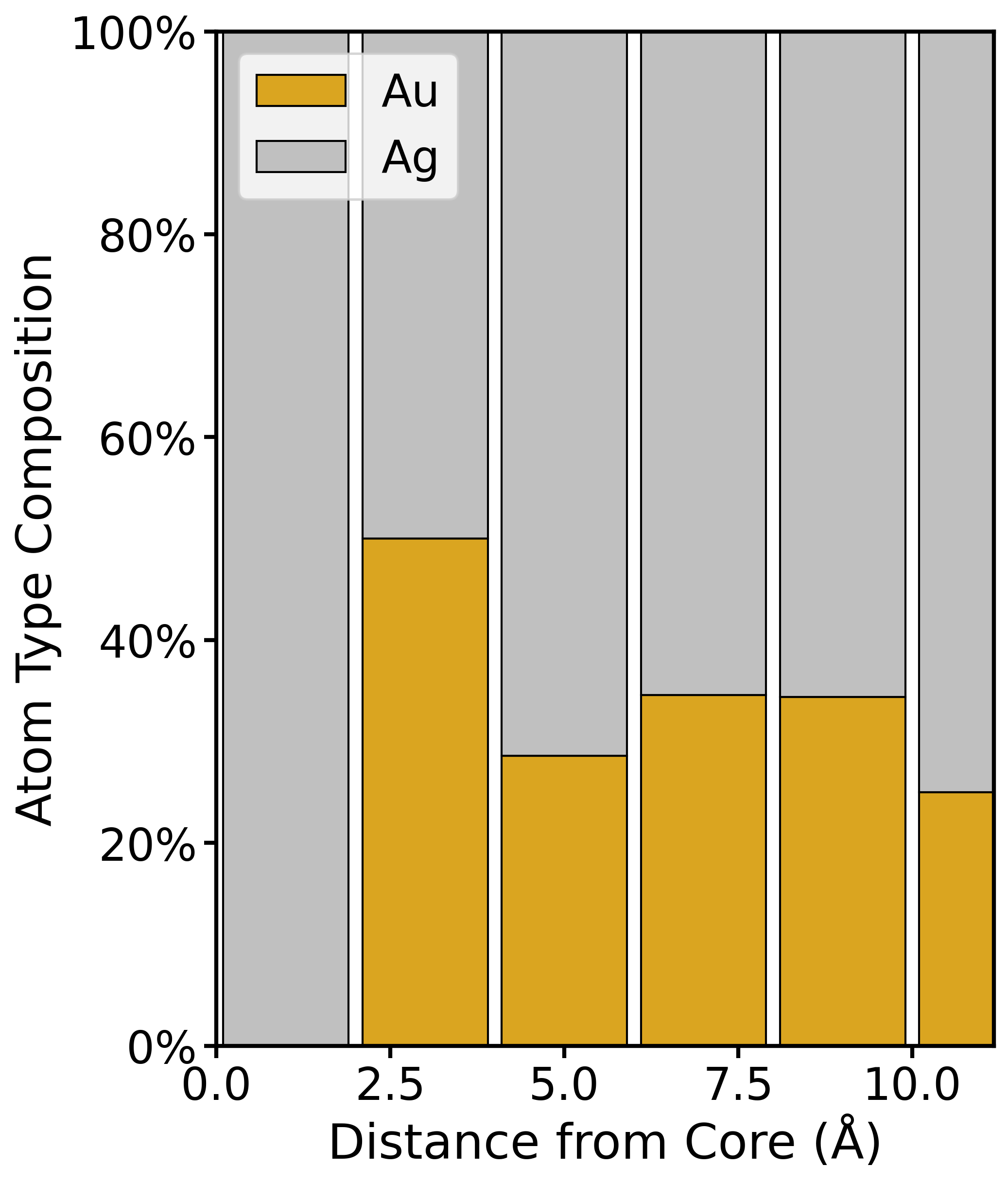}\hfill
  \includegraphics[width=0.44\linewidth]{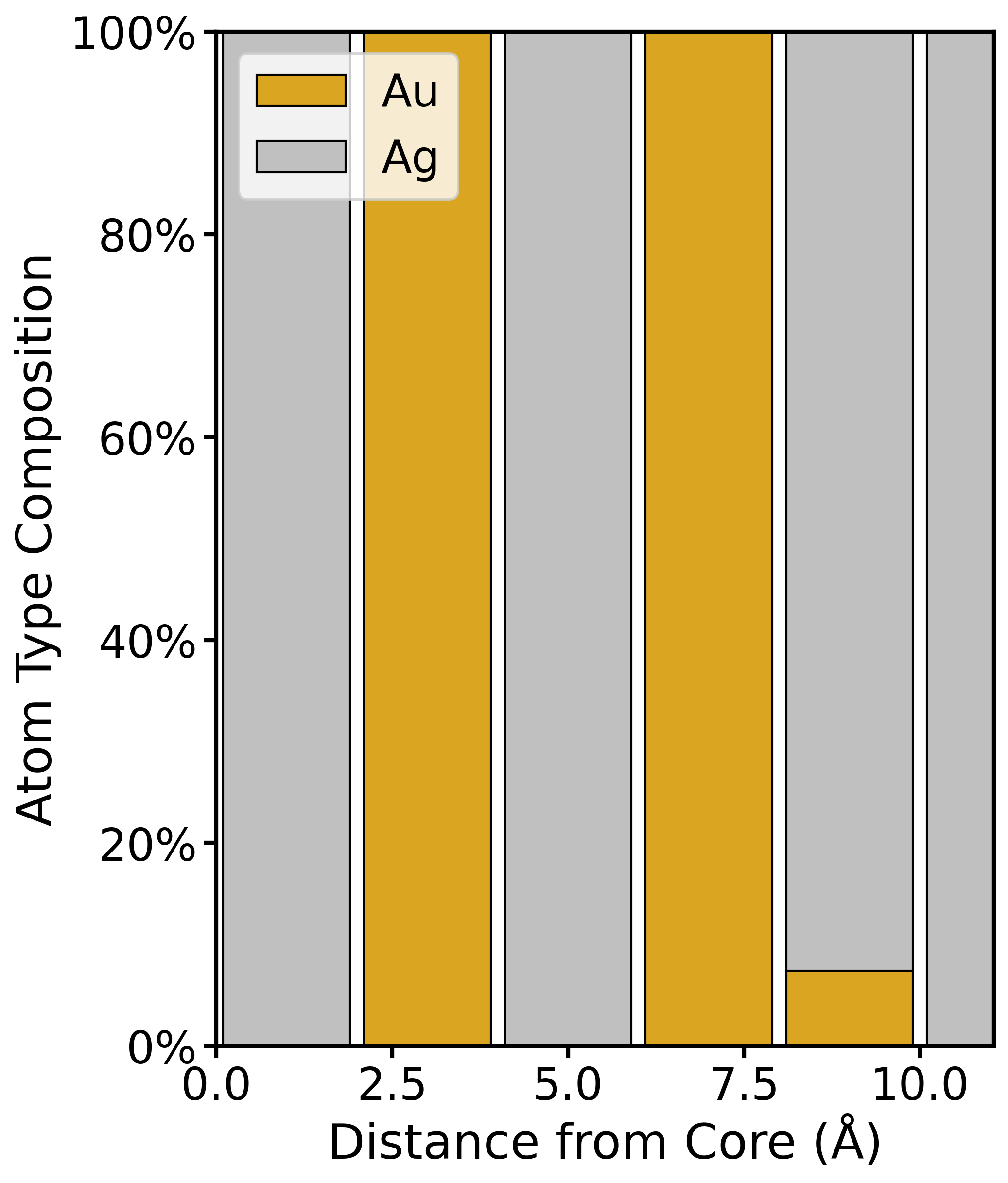}
  \subcaption{3rd best ($52.916388 \ \mathrm{eV} \rightarrow  49.276264 \ \mathrm{eV}$)}
\end{subfigure}\hfill
\begin{subfigure}[t]{0.49\textwidth}
  \centering
  \includegraphics[width=0.44\linewidth]{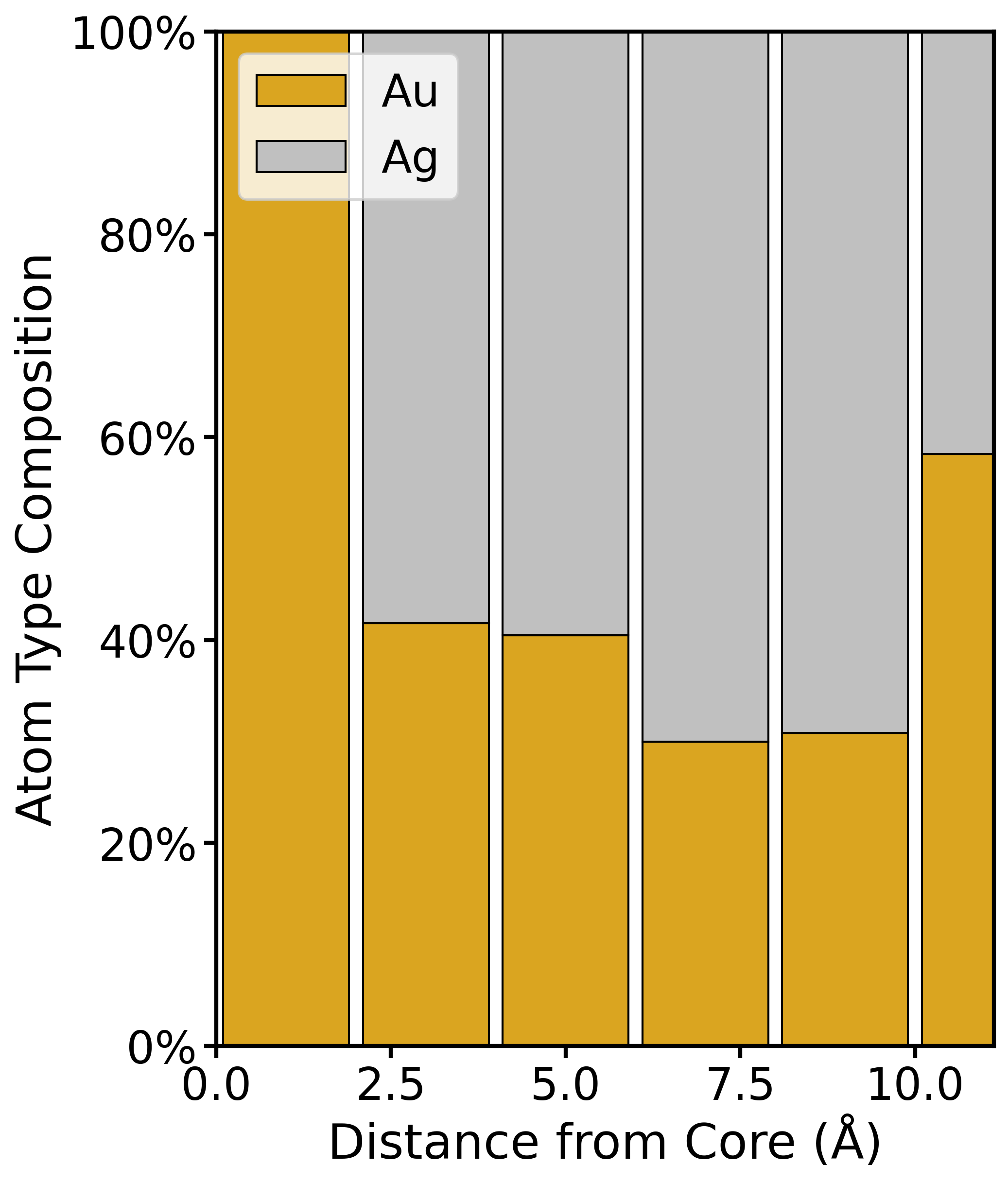}\hfill
  \includegraphics[width=0.44\linewidth]{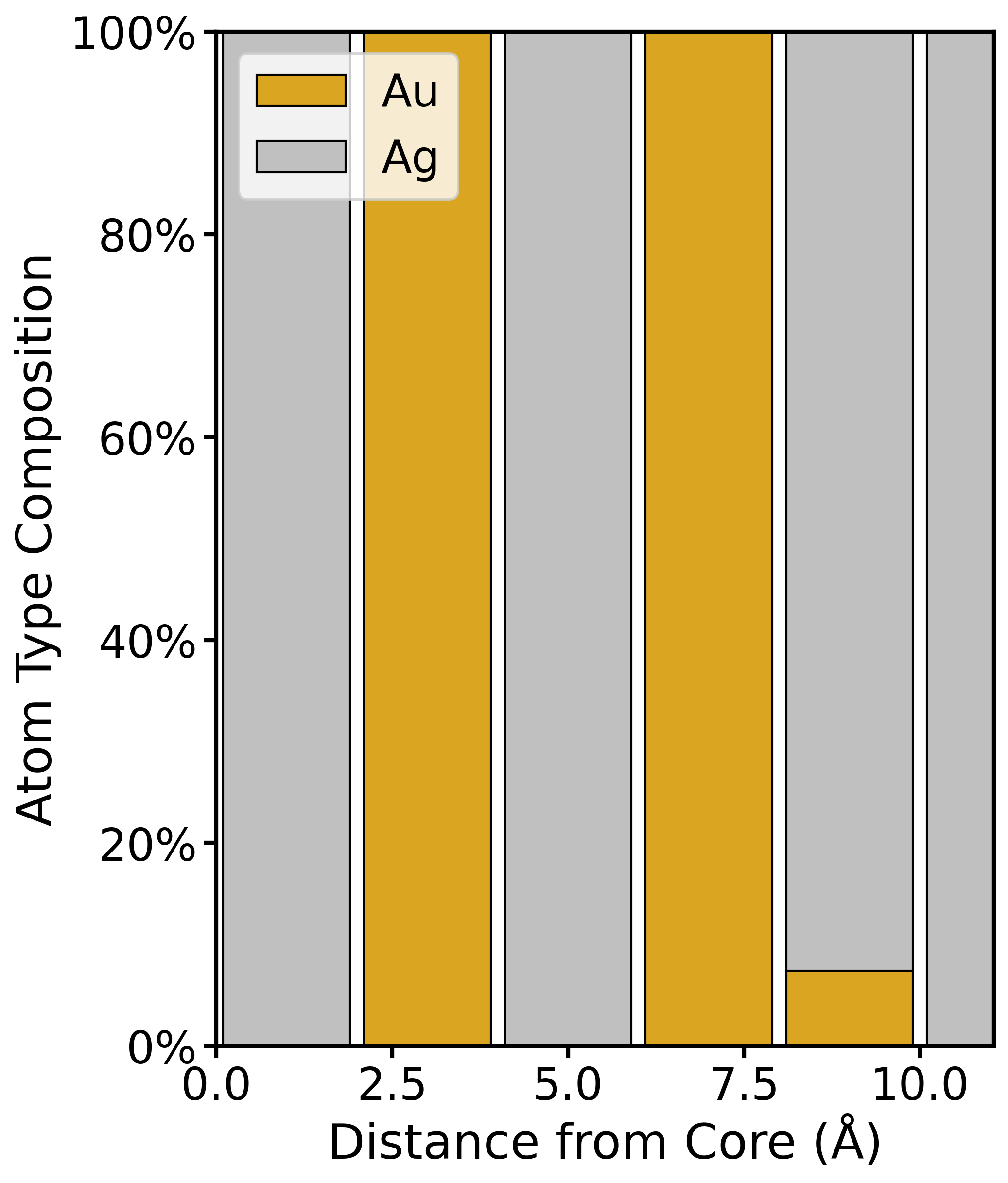}
  \subcaption{4th best ($52.733396 \ \mathrm{eV} \rightarrow  49.276268 \ \mathrm{eV}$)}
\end{subfigure}
% ---------- Row 3 ----------
\begin{subfigure}[t]{0.49\textwidth}
  \centering
  \includegraphics[width=0.44\linewidth]{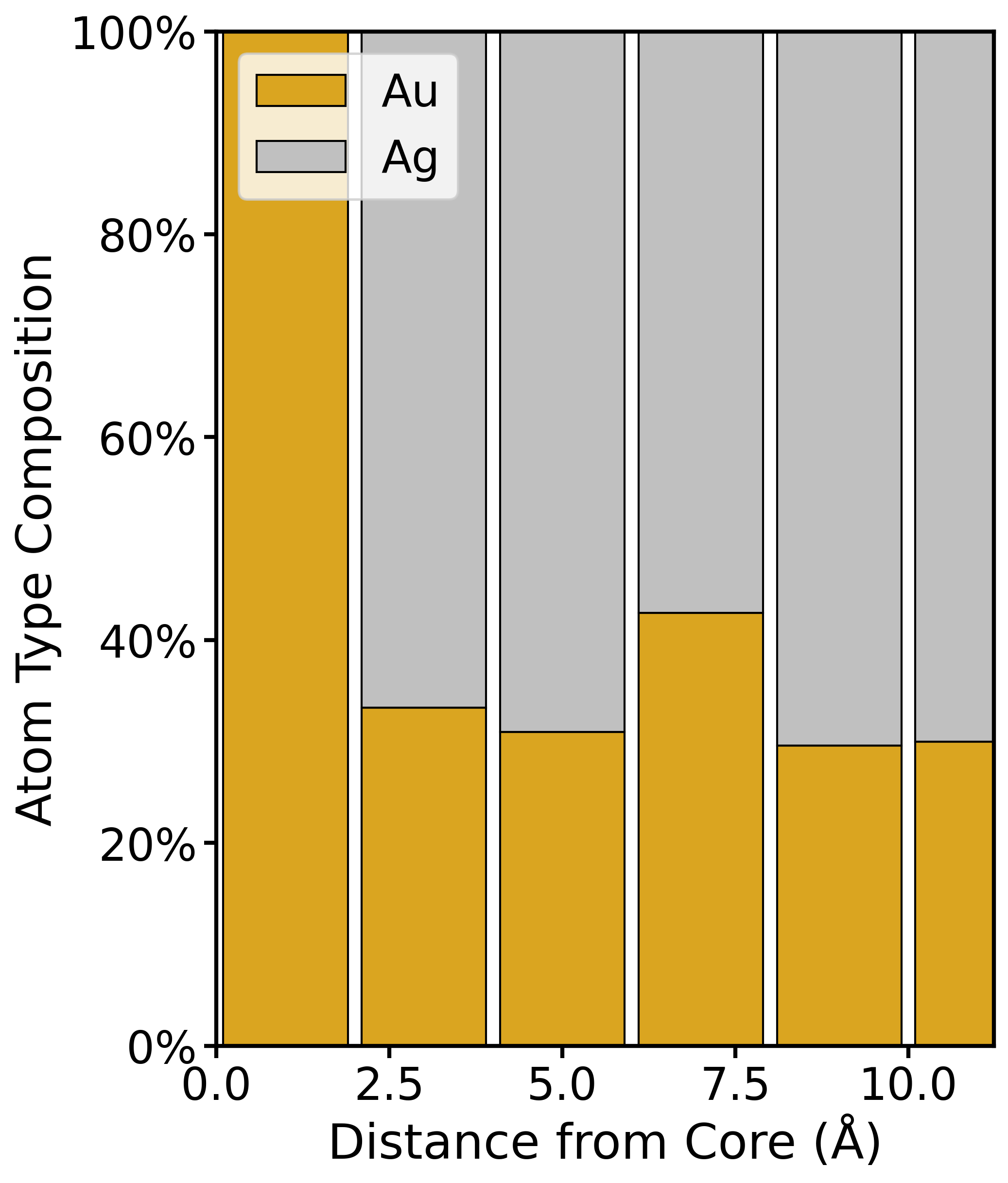}\hfill
  \includegraphics[width=0.44\linewidth]{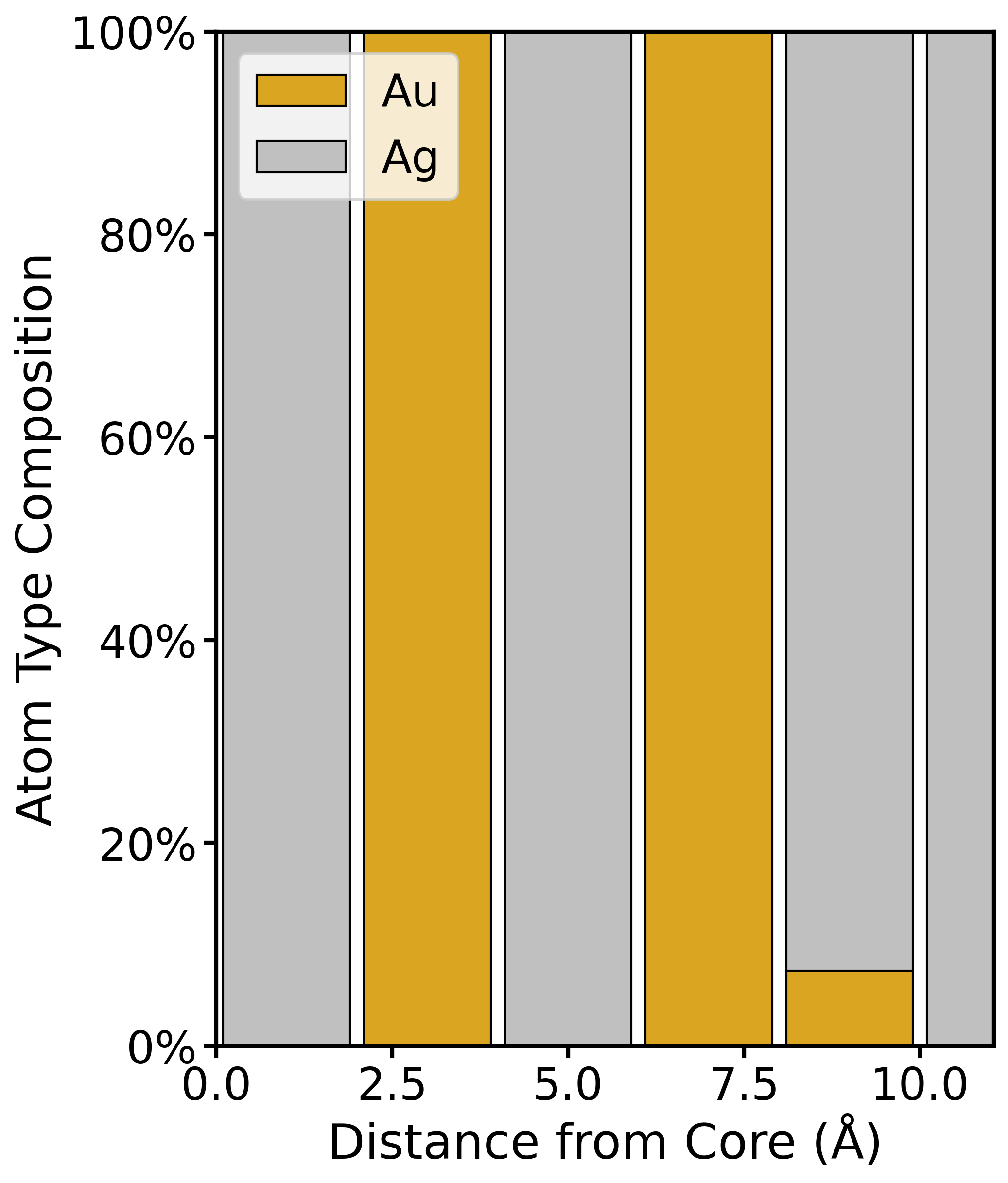}
  \subcaption{5th best ($52.663337 \ \mathrm{eV} \rightarrow  49.276269 \ \mathrm{eV}$)}
\end{subfigure}\hfill
\begin{subfigure}[t]{0.49\textwidth}
  \centering
  \includegraphics[width=0.44\linewidth]{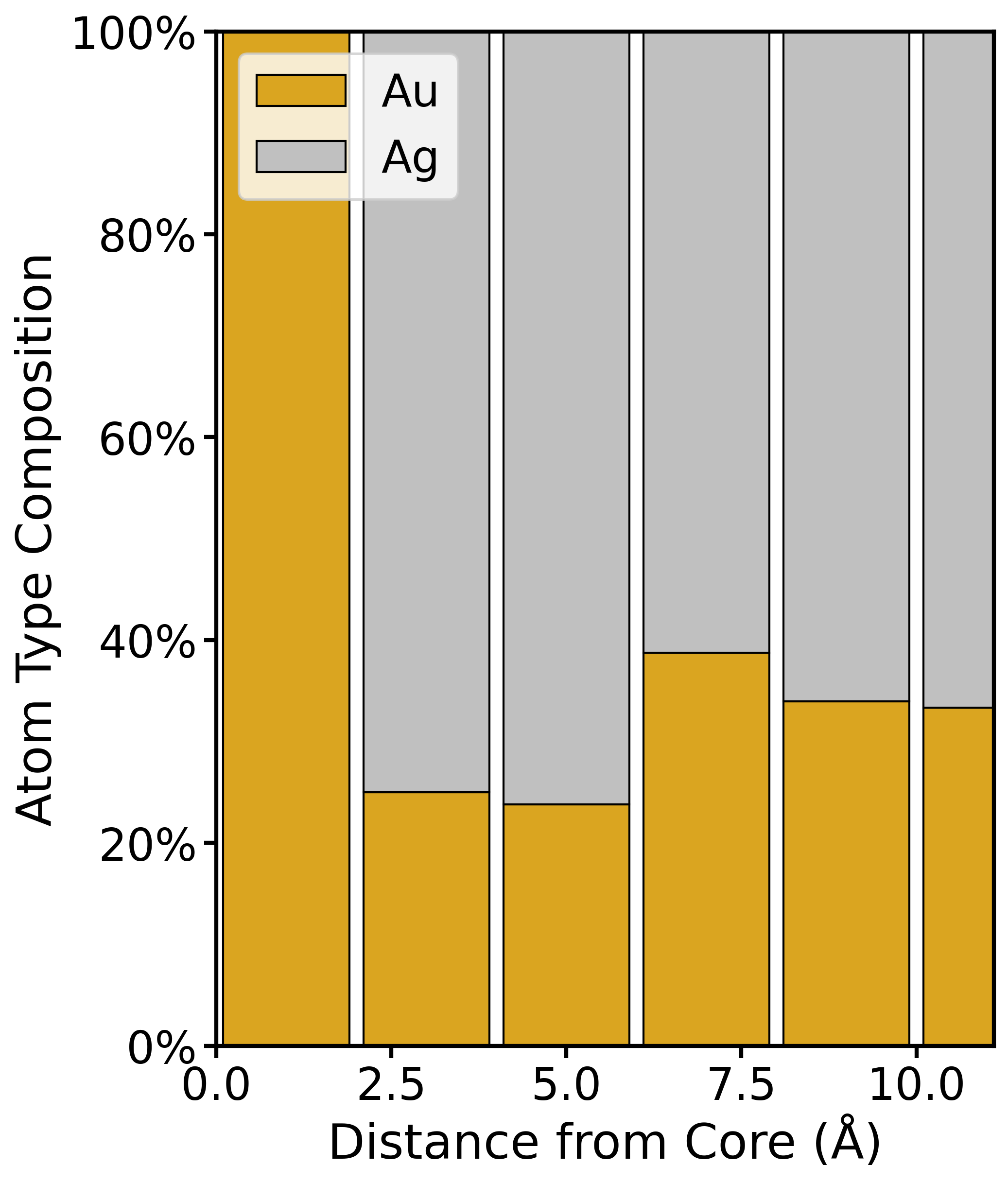}\hfill
  \includegraphics[width=0.44\linewidth]{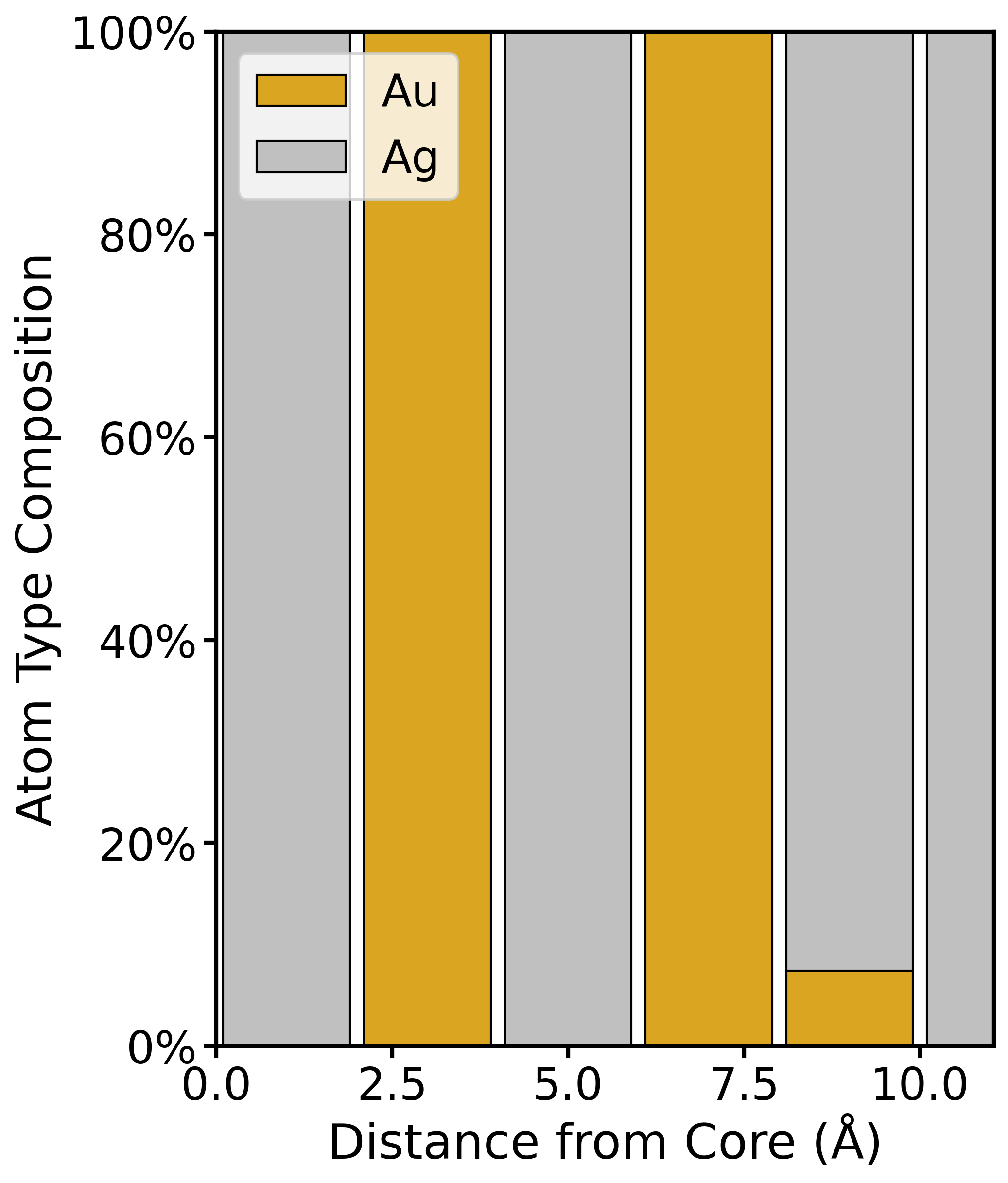}
  \subcaption{6th best ($52.464389  \ \mathrm{eV} \rightarrow  49.276269 \ \mathrm{eV}$)}
\end{subfigure}
% ---------- Row 4 ----------
\begin{subfigure}[t]{0.49\textwidth}
  \centering
  \includegraphics[width=0.44\linewidth]{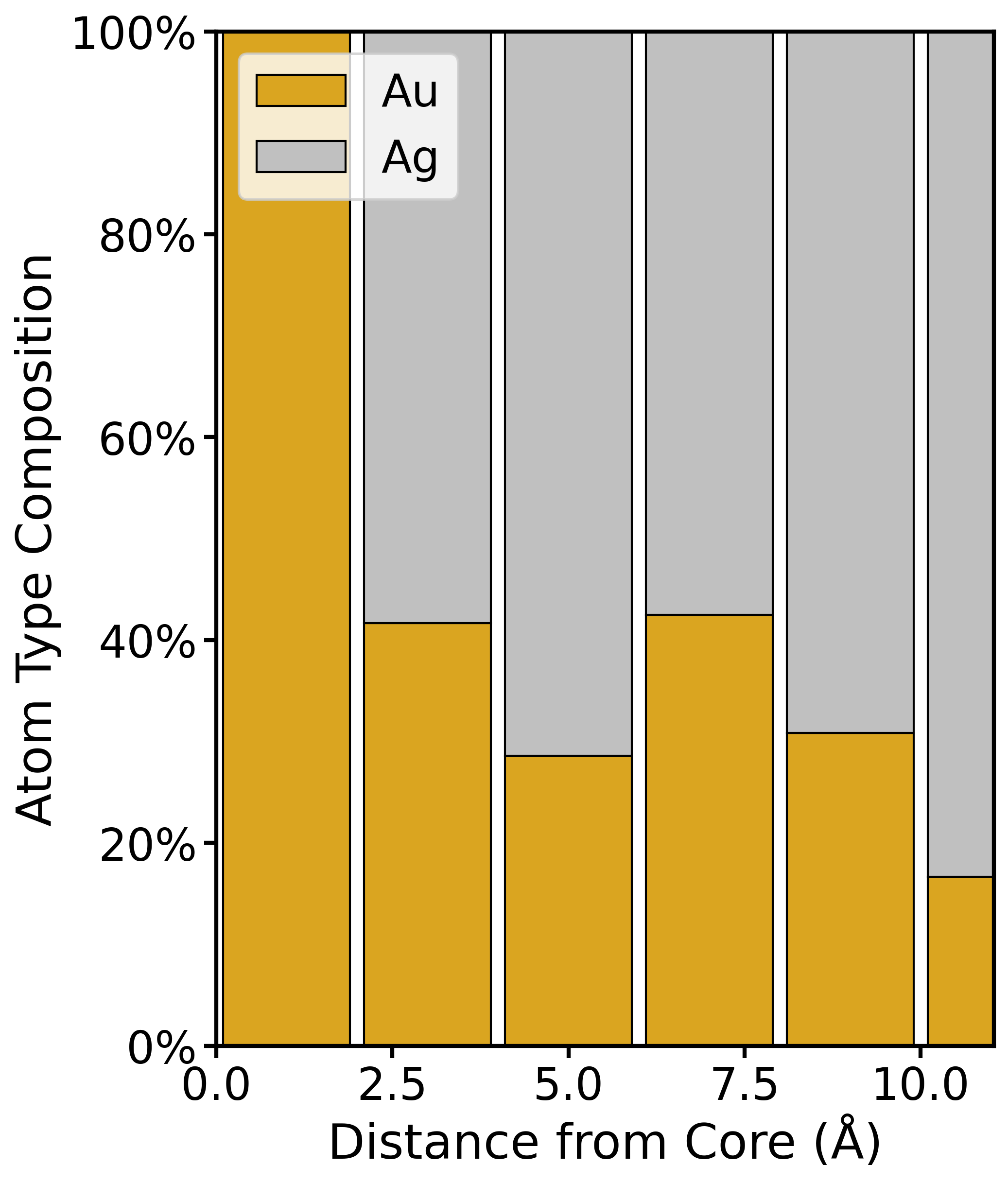}\hfill
  \includegraphics[width=0.44\linewidth]{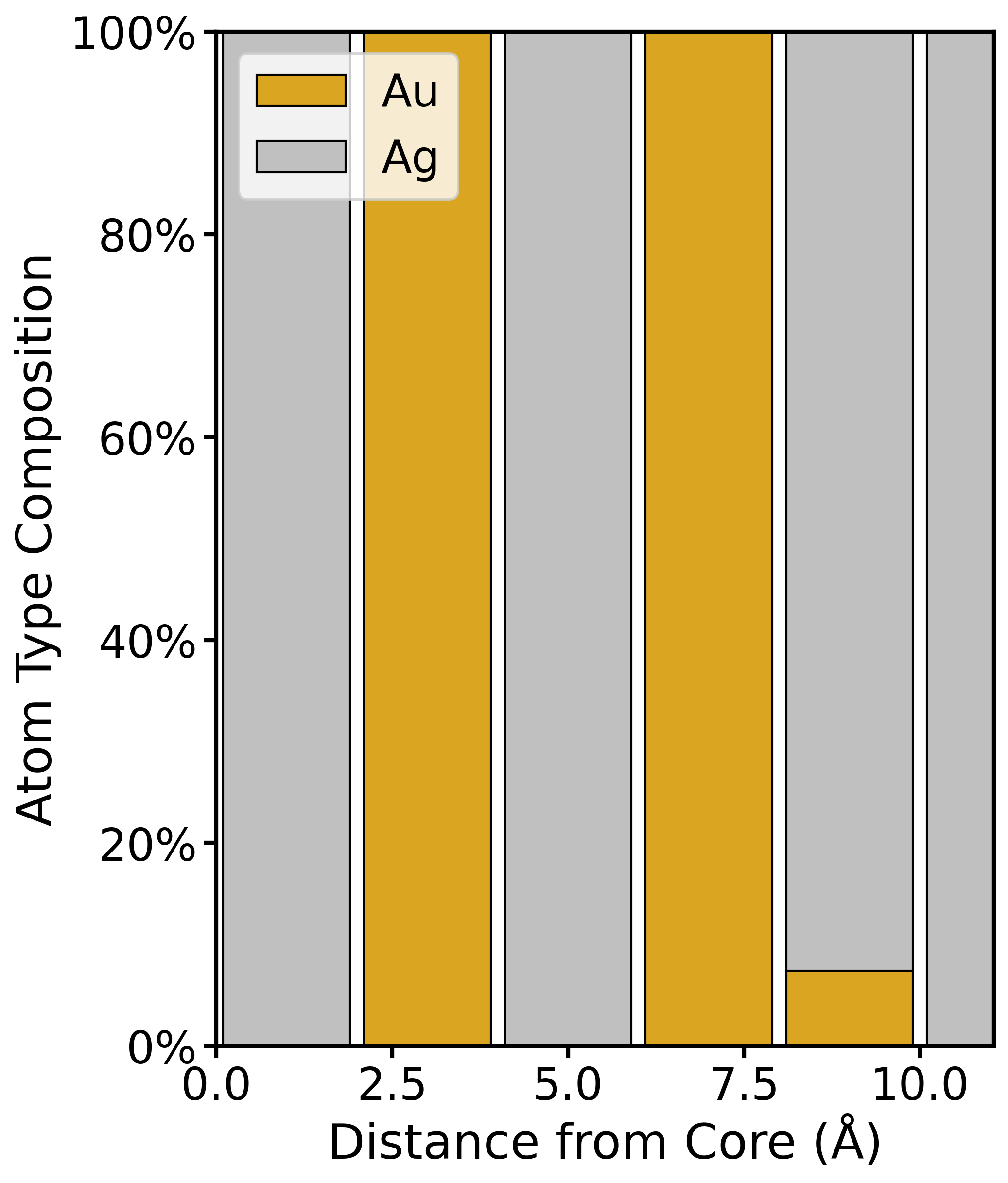}
  \subcaption{7th best ($53.017406   \ \mathrm{eV} \rightarrow  49.276283 \ \mathrm{eV}$)}
\end{subfigure}\hfill
\begin{subfigure}[t]{0.49\textwidth}
  \centering
  \includegraphics[width=0.44\linewidth]{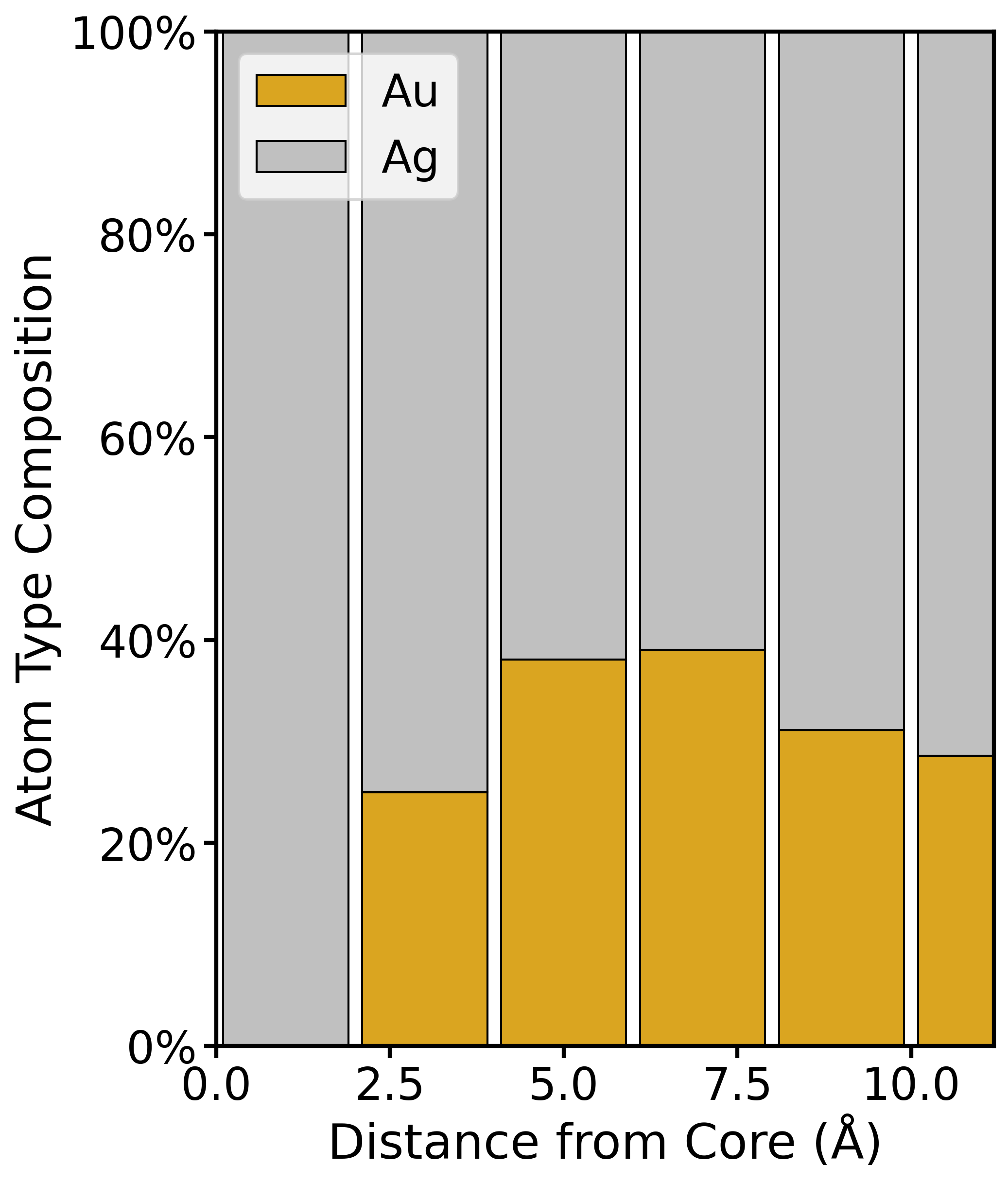}\hfill
  \includegraphics[width=0.44\linewidth]{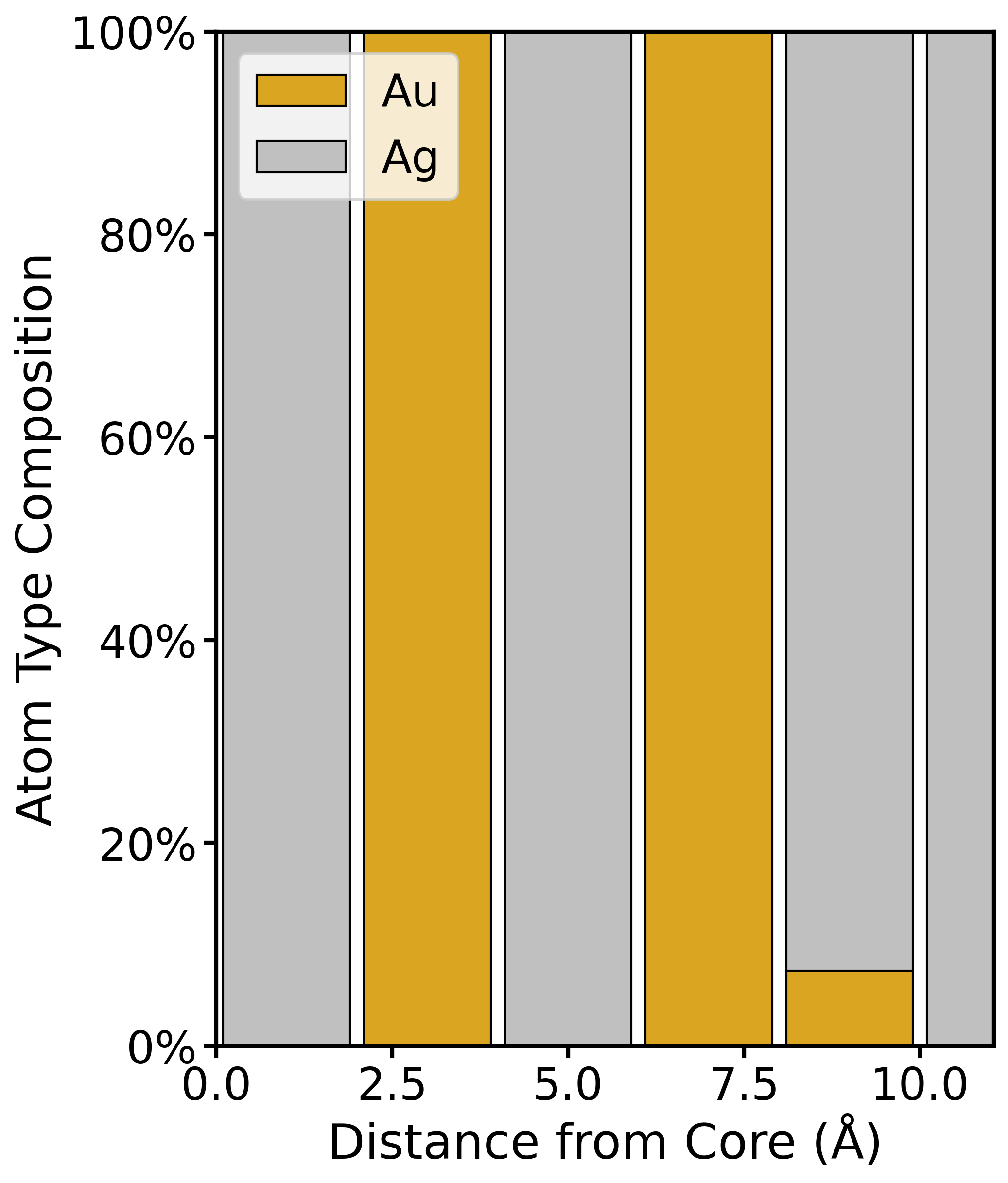}
  \subcaption{8th best ($53.037049 \ \mathrm{eV} \rightarrow  49.308132 \ \mathrm{eV}$)}
\end{subfigure}
\caption{Elemental radial distribution function (E-RDF) plots and energies for eight initialised and final structures of icosahedral \ch{Ag205Au104}. The panels \textbf{(a)-(h)} are ordered from lowest to highest.}
\label{fig:top8-panels}
\end{figure*}
To demonstrate the robustness of the policy-based optimization of elemental ordering against random initialization, we analyze eight final structures optimized by the agent from distinct random initializations of the icosahedral \ch{Ag205Au104} (Figure \ref{fig:top8-panels}). \ch{Ag205Au104} has a verifiably optimal "onion-shell"-like ground state ordering under the EMT potential~\citep{larsen2018rich}. This structure, which was also found by the agent (see Figure \ref{subfig:ih3}), has alternating layers of \ch{Ag} and \ch{Au}, starting with a central \ch{Ag} atom, and thus has an easily recognizable signature in the elemental radial distribution function (E-RDF) plots. In Figure \ref{fig:top8-panels}a-h, we order the eight solutions found by the agent from different random initializations from the lowest final energy (best) (Figure \ref{fig:top8-panels}a, which is that of the configuration in Figure \ref{subfig:ih3}) to the highest final energy (worst). The E-RDF for the randomized initial configurations (left) and for the agent's final structure (right) show that in all cases the agent recovers the shell structure despite varied initialisations, and the energy gaps between Figure \ref{fig:top8-panels}a and the others are negligible ($\approx10^{-5} \ \mathrm{meV/atom}$). The sole outlier (the worst structure, Figure \ref{fig:top8-panels}h) reflects sub-optimal relaxation, and the energy delta is $\approx0.1 \mathrm{meV/atom}$ to Figure \ref{fig:top8-panels}a, yet the elemental ordering and E-RDF remain visually identical.
\textcolor{black}{For the nanoparticle considered here, the RL-optimized structures recover the correct elemental occupation of the outer-shell sites, so the catalytically relevant surface motifs are consistent with those of the corresponding reference ground-state structures. }

\paragraph{Computational cost} \textcolor{black}{The computational cost during both training and deployment comes mainly from the number of swap-relax operations and the typical number of L-BFGS steps per relaxation. The overall cost is dominated by these repeated relaxations rather than by policy inference. In the present setup, intermediate relaxations are capped at 100 L-BFGS steps and typically require about 15–20 steps in practice. For the 309-atom training case, the nominal budget corresponds to 7800 training structures with a horizon of 309 swap-relax steps per structure, with 2.4 million swap-relax operations during training (Figure \ref{fig:training_curves}). However, this setup was not computationally optimized. For binary nanoparticles, the useful horizon during training and deployment should in principle, be limited by the number of atoms of the smaller stoichiometric component, which for the most balanced 309-atom system \ch{Ag155Au154}, would imply a maximum useful horizon of about 154 and an average useful horizon of about 77 when considering all compositions. During deployment, we observe that the structures converge on optimal ordering much earlier than the full horizon length N, e.g. after ~165 swap-relax steps (Figure \ref{fig:full_traj_snaps}). Considering that the number of swaps to achieve correct ordering during deployment is roughly 1/4 of the horizon we employ now (Figure \ref{fig:full_traj_snaps}), the horizon could be shortened, thus reducing the training cost substantially, potentially to \textasciitilde100,000 swap-relax operations with a more optimized protocol. During training we observe that the agent only needs roughly 1/6 of the current horizon to achieve correct ordering.}

\subsection{Policy generalisation allows extrapolation to different-sized NP (Experiment-2)}
\begin{figure}[!ht]
    \centering
    \includegraphics[width=0.99\linewidth]{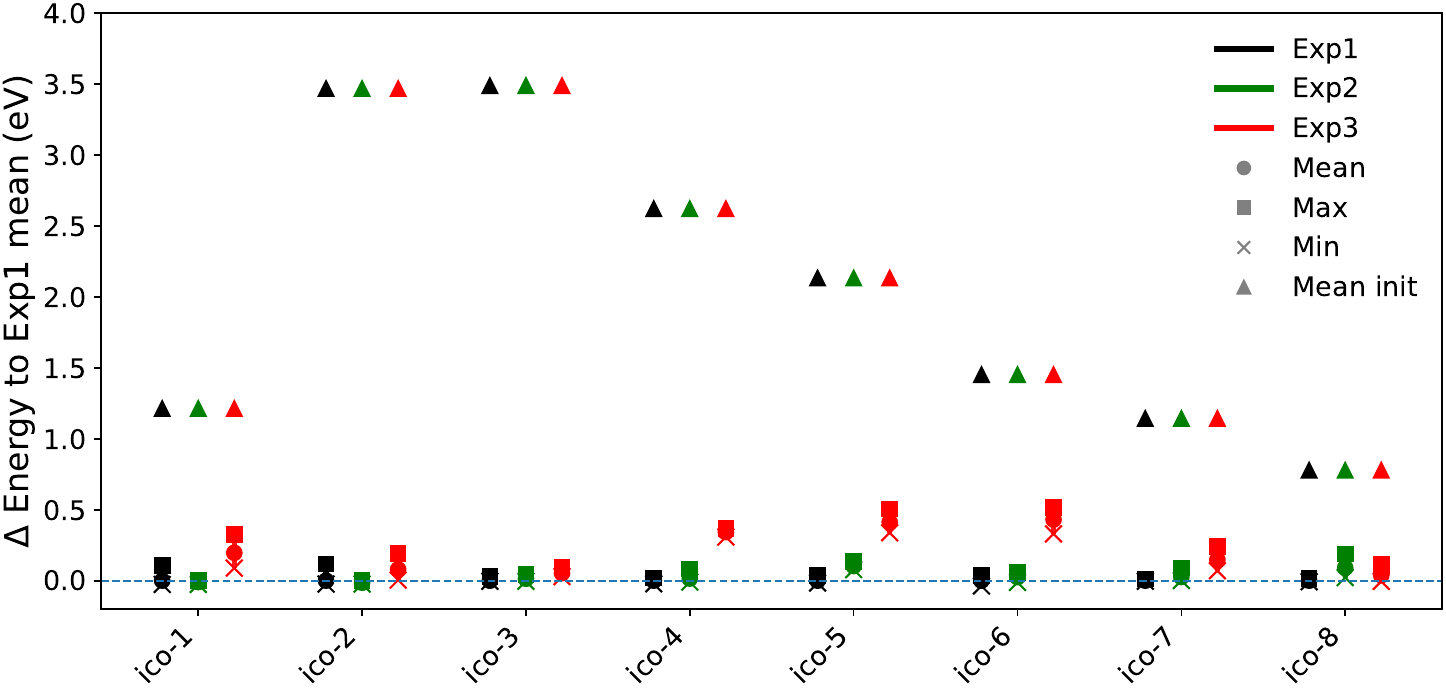}
    \caption{For all eight 309 atom test systems (ico-1 to ico-8), the summary of energies is shown as \(\Delta E\) relative to the average energy of the final optimised structures from eight runs with policy from experiment 1 (dashed line). For each system, we depict the mean energy of the randomly initialised structure; the mean, maximum and minimum energies after optimisation of the structures from all runs. The dashed line at 0 denotes the baseline mean optimised structure energy from experiment 1. Optimisation runs from experiment 2 remain consistent and close to the results obtained in experiment 1. Experiment 3 results show that the policy trained with NPs of multiple element combinations fails to resolve the lowest energy structures in the size extrapolation situation. e.g. for ico-5 and ico-6 the outcomes from different optimisation runs are consistently more than 0.3eV higher in energy than structures found in experiment 1.}
    \label{fig:delta_overlay}
\end{figure}

Extending the success of the model in generalizing over alloy compositions, the next step towards achieving a universal alloy NP structure solver is size generalization and extrapolation capabilities of the policy to unseen-sized NP motifs. 
\paragraph{Training} To test whether the agent’s policy can learn size-invariant ordering rules for icosahedral NP, we train on \ch{Ag_{X}Au_{N-X}} nanoparticles with $N\in\{55,147,561\}$ while \emph{excluding} $N=309$ NPs during the training process. We then evaluate the agent's ability to resolve the lowest energy structure of 309-atom NPs. For consistency of comparison, we check the structures for the same eight Ag-Au compositions (ico-1 to ico-8 in Figure \ref{fig:delta_overlay}) as described in the previous section taken from \citet{larsen2018rich}. 

\paragraph{Results} Comparing the energies of 309 atom NPs (ico-1 to ico-8) optimized with the policy from experiment 1 (trained with 309 atom NPs) against the same NPs optimized with the policy from this experiment (never seen 309 atom NPs during training) provides a benchmark of the model's capability in size extrapolation. Figure~\ref{fig:delta_overlay} shows that deploying the policy for unseen sized NP optimization leads to a slight increase in the energy of the most stable structure found in some cases. The mean $\Delta E$ between the minimum energy structures from these two policies across systems is $\approx 0.021$~eV (see Table \ref{tab:exp_stats}). The robustness of finding the lowest energy structure when optimizing from different starting points is also preserved when extrapolating to 309 atom NP, similar to experiment 1. These results indicate that the learned policy can be reliably transferred across icosahedral shell counts when the alloying elements are consistent.

\begin{table}[!ht]
\centering
\small
\setlength{\tabcolsep}{5pt}
\begin{tabular}{lrrrrrrrr}
\toprule
\textbf{System} & \textbf{E2 $\Delta$Avg} & \textbf{E2 $\Delta$Best} & \textbf{E2 $\Delta$Worst} & \textbf{E3 $\Delta$Avg} & \textbf{E3 $\Delta$Best} & \textbf{E3 $\Delta$Worst} & \textbf{Exp1 spread} \\
\midrule
ico-1 & -0.010 &  0.000 & -0.106 &  0.196 &  0.115 &  0.222 & 0.133 \\
ico-2 & -0.014 & -0.000 & -0.118 &  0.079 &  0.028 &  0.073 & 0.141 \\
ico-3 &  0.014 &  0.000 &  0.015 &  0.053 &  0.033 &  0.065 & 0.032 \\
ico-4 &  0.015 &  0.012 &  0.063 &  0.342 &  0.330 &  0.351 & 0.040 \\
ico-5 &  0.103 &  0.095 &  0.101 &  0.412 &  0.355 &  0.468 & 0.053 \\
ico-6 &  0.025 &  0.025 &  0.024 &  0.431 &  0.367 &  0.480 & 0.073 \\
ico-7 &  0.026 &  0.005 &  0.077 &  0.148 &  0.074 &  0.230 & 0.016 \\
ico-8 &  0.091 &  0.029 &  0.172 &  0.037 &  0.002 &  0.099 & 0.024 \\
\midrule
\textbf{Mean} & \textbf{0.031} & \textbf{0.021} & \textbf{0.028} & \textbf{0.212} & \textbf{0.163} & \textbf{0.248} & \textbf{0.064} \\
\bottomrule
\vspace*{0.2cm}
\end{tabular}
\caption{For each of the eight test 309 atom alloy nanoparticles, the average, lowest and highest energy(eV) after optimisation in experiment 2 (E2) and experiment 3 (E3) is reported relative to the average, lowest and highest energy after optimisation in experiment 1.  The maximum to minimum energy range (Exp1 spread) of the final state indicates intrinsic variability within multiple runs for a specific NP case (ico-1 to ico-8).}
\label{tab:exp_stats}
\end{table}
\subsection{Limited effectiveness in size extrapolation for cross-chemistry generalised policy (Experiment-3)}
A model that can generalize among many different alloy elements and extrapolate to different sized NP can essentially be used to build foundational NP structures solvers that can be deployed across all elements and NP sizes. 

\paragraph{Training}To test the possibility of achieving multi-element generalization and size extrapolation jointly, we added a second bimetallic chemistry during training and checked for robustness and transferability. We trained on both Ag--Au and Pt--Ni for NP sized $N\in\{55,147,561\}$, while \textit{excluding} $N=309$, and the policy was evaluated on the eight Ag--Au 309-atom NP as in the previous two experiments. \textcolor{black}{We show training curves in Figure \ref{fig:training_curves_exp3} in Appendix \ref{app:training_curves}.}
\paragraph{Results}
Introducing Pt-Ni and Ag-Au together in training makes the policy less reliable in finding the optimal structure of unseen 309-atom NPs. Energies of the final optimized structure proposed by this policy trend are noticeably higher than those of experiments 1 or 2 (Figure ~\ref{fig:delta_overlay}). On average, the mean energy of optimized structures increases by $\approx 0.21$~eV between systems, with the largest deviations observed for \textit{ico-5} and \textit{ico-6}. Even considering the most stable optimized structure from each system, the optimized energy increases by $\approx 0.16$~eV. Some systems remain less affected (e.g., \textit{ico-8}), but the optimized NP energy is consistently worse than in experiment 1 and is therefore unreliable for optimization of the NP structure. This points to a chemistry-induced distribution shift: mixing chemistries with distinct ordering energetics biases the policy away from Ag-Au motifs that were optimal in experiment 1. \textcolor{black}{For PtNi nanoparticles, the jointly trained policy recovers chemically sensible orderings that agree qualitatively with prior literature \citep{cardona2023structural}, in particular the Pt-rich outer shell, although some discrepancies remain in the inner-shell arrangement. In particular, the RL framework reproduces key characteristics of the \ch{Ni113Pt196} and \ch{Ni419Pt50} nanoparticles reported previously \citep{han2022unfolding}, especially for the outer shells, where the Pt-rich surface structure is recovered correctly. The main discrepancies appear in the inner shells (Figure \ref{fig:PtNi})}. 

\subsection{\textcolor{black}{Robustness of inference}}
\textcolor{black}{With Figure~\ref{fig:delta_overlay} we intended to assess the robustness and consistency of the three training setups across repeated optimizations from random initial chemical orderings, rather than only their single best-case performance. For this reason, the final energies are plotted relative to the average energy of the eight optimized runs for each experiment. In experiments 1 and 2, the optimized structures converge essentially to the same final ordering, and the small residual energy differences arise mainly from minor numerical variations in the L-BFGS relaxation (\textasciitilde0.01 meV) rather than from physically distinct chemical orderings or structures. By contrast, experiment 3 exhibits a larger spread in final energies and more frequent failure to recover the lowest-energy ordering, indicating reduced robustness and consistency in the multi-chemistry setting. A comparison to the lowest-energy structure is the more natural metric for benchmarking a global optimizer; this coincides with the baseline marker. For experiments 1 and 2, the near-identical final energies across repeated runs already indicate that the final structure reaches the global optimum but for experiment 3, a cumulative success-rate analysis may be more informative if there are lower energy variants encountered during the swap-relax steps.}
\section{\textcolor{black}{Discussion}}
In this work, we have shown that, utilizing a molecular graph representation, an RL agent can reproduce provably optimal ground state nanoparticle orderings. The agent finds these orderings for varying compositions and can converge on identical optimal solutions from differently ordered initialisations of the same chemical composition. The proposed architecture and training scheme allow for successful deployment of the agent for NPs that have a different size than training NPs, but this size extrapolation is not robust if multiple different element combinations must be tackled with the same policy. 

In the future, this RL framework needs to be further developed to enable generalization to larger combinatorial spaces, encompassing more atoms, atomic species, and other nanoparticle motifs and morphologies. \textcolor{black}{The wider chemical space of NPs remain beyond the reach of traditional methods such as genetic algorithms, Monte Carlo, basin hopping and minimum integer programming. Compared with these classical optimization approaches the present RL method has a higher upfront training cost but offers the possibility of cost amortization across many related optimization problems. Classical methods typically require a fresh search for each new nanoparticle size or composition, whereas a trained RL policy can be reused and can reach low-energy orderings using relatively few additional swap-relax steps during deployment. We therefore do not claim that the present approach is a universal replacement for established methods in single-instance optimization, but rather that it becomes attractive when a broad family of related chemical-ordering problems must be solved repeatedly.}  \textcolor{black}{However, it is worth noting that RL methods can also benefit from efficient search strategies developed for these classical approaches. In particular, symmetry-constrained search has proven highly effective in genetic-algorithm-based nanoparticle optimization by drastically reducing the accessible ordering space while retaining chemically relevant low-energy motifs~\cite{han2022unfolding}. This suggests that symmetry-constrained RL could also be a promising approach to improve efficiency for highly symmetric nanoparticle motifs, for example by restricting actions to symmetry-inequivalent sites or groups. } The search cost scales superlinearly while the objective landscape grows more complex. With RL, we hope to amortize exploration by learning transferable action priors and value estimates that can guide sampling toward high-yield regions with potentially fewer energy calls than naive search. 

On the implementation side, the currently used frozen ORB-v3 backbone is a pragmatic choice, but its pretraining emphasizes crystalline bulk datasets (e.g. MPtraj, Alexandria) rather than nanoparticles. Although the architecture supports non-periodic systems, nanoparticle or surface-specific data are missing in the training set. Future iterations of our agent could thus benefit from, e.g., nanoparticle-trained interatomic potentials and/or from performing end-to-end fine-tuning of the encoder jointly with the policy/value heads on NP data. A complementary direction is to replace expensive per-step relaxations with learned relaxations. 
\textcolor{black}{The present model extends earlier on-lattice periodic alloy ordering RL framework to nanoparticle alloy ordering, in addition also implicitly accounting for the effect of local distortions through structure relaxation during training. In principle, one might consider training a policy that learns both atomic swaps and explicitly performs lattice distortions toward low-energy structures, but this would introduce a much larger continuous action space associated with the displacement of all atomic positions and is therefore left for future work. Although the present environment is restricted to pair swaps to maintain a tractable action space and a well-defined per-step reward, the framework can in principle be extended to multi-atom moves through cyclic k-atom permutations, learned multi-swap actions, or hierarchical action policies, at the cost of a substantially larger action space and more difficult credit assignment.}
In Figure \ref{fig:results}, we observed that in the final proposed structure from RL, a few atoms can be placed sub-optimally compared to the ideal solution. One reason for this could be that we currently have no mechanism to prevent energy-increasing moves late in the episode because same-species swaps are explicitly prohibited by masking. In practice, the agent thus currently learns to keep swapping atoms back and forth between the same positions in an attempt to maintain an optimal structure. This is not ideal, so an improvement to our method could be to implement a mechanism to avoid forcing non-productive moves. Accordingly, we will experiment with the removal of the same-species mask and introduce either a no-operation/stop action or a variable-length horizon. The latter could be implemented by introducing a per-step penalty to discourage unnecessary swaps. Together these changes might let the agent preserve the already optimal structures, anticipate relaxation effects, and generalize better as we expand our model to include more elements, nanoparticle sizes, and motifs.

\section{Data availability statement}
The codebase for this work is openly available at \url{https://github.com/Jotels/medusa}.

\section{Acknowledgments}
We received funding from DNRF for the Pioneer Center for Accelerating P2X Materials Discovery (CAPeX) grant number P3, from the Det Frie Forskningsråd for Project “Data-driven quest for TWh scalable Na-ion battery (TeraBatt)” (2035-00232B) and “Autonomous agents of Discovery for earth-Abundant Na-ion battery cathodes (ADANA)” (3164-00297B). We also thank the Novo Nordisk Foundation Data Science Research Infrastructure 2022 Grant: A high-performance computing infrastructure for data-driven research on sustainable energy materials, Grant no. NNF22OC0078009.

\newpage
\appendix
\section{Algorithm details} \label{appendix:algorithm}
Below, we detail the algorithm and implementation choices we have made. A graphical overview of the method is provided in Figure \ref{fig:agau309-ppo}.
\subsection{MDP and rewards}\label{app:mdp}
Each episode is a finite-horizon MDP
\[
\mathcal{M}=(\mathcal{S},\mathcal{A},P,r,H,\gamma),
\]
over a 309-atom icosahedral nanoparticle with a fixed random composition $\ch{Ag_{X}Au_{309-X}}$ and randomly initialized ordering/positions. Here
$\mathcal{S}$ is the state space, $\mathcal{A}$ the discrete action space,
$P:\mathcal{S}\times\mathcal{A}\to\Delta(\mathcal{S})$ is the transition kernel
$P(s'|s,a)$,
$r:\mathcal{S}\times\mathcal{A}\times\mathcal{S}\to\mathbb{R}$ is the reward function,
$H\in\mathbb{N}$ is the episode length, and $\gamma\in[0,1]$ is the discount.
In our environment, the transition $s_{t+1}\sim P(\cdot|s_t,a_t)$ is the \emph{swap+relaxation} step (see Figure~\ref{fig:step_fig}). In our setup, this transition is approximately deterministic.
$E_t$ is the potential energy after step $t$, and $E_0$ is the initial energy of the randomized configuration of the nanoparticle.
With per-step relaxation, we define the instantaneous reward
\begin{equation}
    r_t \;=\; E_{t-1}-E_t,\quad t=1,\dots,H.
\end{equation}
Because we do not use a discount factor for our rewards, \textit{i.e.} $\gamma=1$, the telescoping property holds exactly for our setup:
\begin{equation}
    G=\sum_{t=0}^{H-1} r_t=\sum_{t=0}^{H-1}\left(E_t-E_{t+1}\right)=E_0-E_1+E_1-E_2+\cdots+E_{T-1}-E_T=E_0-E_H,
    \label{eq:telescoping}
\end{equation}
so maximizing undiscounted return is equivalent to minimizing final energy. We interface with the nanoparticle using the Atomic Simulation Environment (ASE)~\citep{larsen2017atomic}, and similarly compute energies in ASE. \\
For these calculations, we use the Effective Medium Theory (EMT) potential implemented in ASE~\citep{jacobsen1987interatomic,jacobsen1996semi}. EMT is a semi-empirical tight-binding/second-moment–type model in which the total energy decomposes into a short-ranged repulsive pair term plus an embedding contribution that depends on a locally accumulated electron density from neighbours. Both contributions are modeled by decaying exponentials with element-specific parameters fitted to reproduce bulk lattice constants, cohesive energies, bulk moduli and vacancy formation energies of the pure elements. Heteronuclear (alloy) interactions follow the standard EMT mixing rules implemented with the same parameter set. These parameters are fitted to reproduce experimental and Density Functional Theory (DFT) data for the elemental metals Ni, Cu, Ag, Au, Pd, Pt and bimetallic alloys~\citep{jacobsen1996semi}. EMT thus captures the correct trends in relative stabilities of different configurations, such as the energetic ordering of alloy structures or cluster isomers~\citep{lysgaard2014genetic}. In practice we use the compiled ASAP implementation, which provides a vectorized neighbor-list kernel in C/C++ with OpenMP parallelism. Since we work with nanoparticles, we disable periodic boundary conditions and add $10$ Å of vacuum so that EMT’s short cutoff does not couple periodic images. \\
To perform local relaxations, we use the line-search L-BFGS optimizer~\citep{liu1989limited} implemented in ASE. We use a maximum-force parameter $f_{\max}$ to decide when to stop the relaxation $\max_i\|\mathbf{F}_i\|_\infty \le f_{\max}$ or alternatively when a step budget $N_{\text{Max}}$ is reached. In practice we use small relaxations at intermediate steps ($f_{\max}=0.01$, $N_{\text{Max}} = 100$) and a stricter final relaxation at $t = H$ ($f_{\max} = 5 \times 10^{-3}$, $N_{\text{Max}} = 1000$). The reward is the energy drop after the \emph{swap+relax} step has taken place.

\subsection{State encoding and factorized policy} \label{app:policy_factor_details}
At time step $t$, the state $s_t$ is the current nanoparticle (symbols and positions), which we encode using a pretrained ORB-v3 encoder~\citep{rhodes2025orbv3atomisticsimulationscale}. This yields per-atom embeddings
$\mathbf{h}_i\in\mathbb{R}^{D_0}$ for atom $i$.
The action is a swap $a_t=(i_t,j_t)$ between an \emph{anchor} index $i_t$ and a \emph{partner} index $j_t$.
We parameterize the joint policy by a factorization
\begin{equation}
\pi_\theta(a_t\mid s_t)
=
\pi_{\theta_a}(i_t\mid s_t)\;
\pi_{\theta_p}(j_t\mid s_t,i_t),
\label{eq:factored_policy}
\end{equation}
with $\theta=(\theta_a,\theta_p)$.

\paragraph{Time embedding.}
We make the policy explicitly time–aware by concatenating a fixed, non-learned positional code of the current step and a scale indicator of the episode length. Let $t\in\{0,\dots,H-1\}$ denote the step index and $H$ the horizon. We define
\[
\psi(t,H)=\big[PE(t),\ \log H\big]\in\mathbb{R}^{d+1},
\]
where $PE(t)\in\mathbb{R}^{d}$ is the sinusoidal code
\[
PE_{2m}(t)=\sin \Big(\frac{t}{10000^{2m/d}}\Big),\qquad
PE_{2m+1}(t)=\cos \Big(\frac{t}{10000^{2m/d}}\Big),\quad m=0,\dots,\tfrac{d}{2}-1.
\]
The inclusion of $\log H$ conditions the policy on the remaining budget. While we use fixed horizons due to the fixed particle size, we aim to use this to promote transfer across tasks with different horizons in future implementations where particle sizes and motifs vary. In implementation, $\psi(t,H)$ is tiled to all atoms so that each per-atom feature vector at time $t$ carries identical temporal context. In theory, this design breaks step-exchange symmetry (action preferences near the end of an episode may differ from those early on).

\paragraph{Anchor head.}
Let $\mathbf{h}_i\in\mathbb{R}^{D_0}$ be the per-atom embedding from the pretrained ORB-v3 encoder for atom $i$. The ORB-v3 encoder also outputs an $\ell = 1$ vector channel $\mathbf{v}_i\in\mathbb{R}^3$, which is a predicted force per atom. In our implementation, we recognize that the ORB-v3 model is not tailored to nanoparticles, and thus the predicted forces $\mathbf{v}_i$ are not used directly to, e.g., relax the structure, but only used as force/dispersion "proxies", whose relation (if any) to the optimal ordering is learned inside the policy networks. To produce the anchor policy distribution, we compress this force proxy to a rotation-invariant scalar $\|\mathbf{v}_i\|_2$ and map it through a small MLP to obtain $\boldsymbol{\chi}_i\in\mathbb{R}^{d_\chi}$. We provide $\boldsymbol{\chi}_i$ to the agent in order to incentivize toward atoms with large local directional signal (high stress/instability), while $\|\mathbf{v}_i\|_2=\|R\mathbf{v}_i\|_2$ for any $R\in\mathrm{SO}(3)$ ensures frame-invariant logits. The time-aware anchor features are
\begin{equation}
    \boldsymbol{\phi}^{(a)}_i(t)=\big[\ \mathbf{h}_i,\ \boldsymbol{\chi}_i,\ \psi(t,H)\ \big]\in\mathbb{R}^{D_0 + d_\chi + (d+1)}.
\end{equation}
An MLP $g_{\theta_a}:\mathbb{R}^{D_0 + d_\chi + (d+1)} \to\mathbb{R}$ produces a scalar logit for each atom,
\begin{equation}
    z^{(a)}_i = g_{\theta_a} \big(\boldsymbol{\phi}^{(a)}_i(t)\big),
\end{equation}
and the anchor distribution over indices $\{1,\dots,N\}$ is the categorical
\begin{equation}
    \pi_{\theta_a}(i\mid s_t)=\mathrm{softmax} \big(z^{(a)}\big)_i
\quad\text{with}\quad
z^{(a)}=\big(z^{(a)}_1,\dots,z^{(a)}_N\big).
\label{eq:anchor_dist}
\end{equation}
We do not rule out any atoms at the anchor stage (all atoms may serve as anchors). Instead, any masking is deferred to the partner head. Computationally, the anchor pass is $O(N)$ and $\psi(t,H)$ is broadcast to all atoms at step $t$.

\paragraph{Partner head (conditional on $i$).}
Given a sampled anchor $i$, we concatenate to each candidate partner atom indexed by $j$ the embedding $\mathbf{h}_j$ and the broadcast anchor embedding $\mathbf{h}_i$. We also augment the partner candidate features with two E(3)-invariant pairwise descriptors: 
\begin{equation}
    d_{ij}=\|x_j-x_i\|_2,\qquad
a_{ij}=\mathbf{v}_i^\top\mathbf{v}_j \ ,
\end{equation}
where $d_{i j}$ is a radial separation and $a_{i j}$ is a force-alignment term derived from the encoder's force predictions. Similar to the anchor features, the relation to the force-alignment term is learned, and the forces are not used directly. These two terms thereby inject a lightweight physics prior, allowing the policy to learn a signed preference (favor or penalize) for near-field swaps conditioned on these descriptors, while keeping the logits frame-invariant. The full partner features are therefore
\begin{equation}
    \boldsymbol{\phi}^{(p)}_{j\mid i}(t)
=
\big[\ \mathbf{h}_j,\ \mathbf{h}_i,\ \bar{\mathbf{e}}_{i\to j},\ d_{ij},\ a_{ij},\ \psi(t,H)\ \big],
\end{equation}
passed through an MLP $g_{\theta_p}$ to produce logits $z^{(p,i)}_j=g_{\theta_p}(\boldsymbol{\phi}^{(p)}_{j\mid i})$.
Feasibility is enforced by a masked categorical
\begin{equation}
    \pi_{\theta_p}(j\mid s_t,i)=\operatorname{softmax} \big(z^{(p,i)}+\mathbf{m}_{s_t,i}\big)_j,\qquad
\mathbf{m}_{s_t,i}(j)=
\begin{cases}
-\infty,& Z_j=Z_i,\\
0,& \text{otherwise,}
\end{cases}
\label{eq:partner_dist}
\end{equation}
which prohibits same-species swaps and (therefore) self-swaps, which has the added effect of reducing the effective action space while preserving correct invariances of the logits.

\paragraph{Critic and value pooling.}
The value function \(V_\phi(s_t)\) shares the time–aware per-atom inputs with the anchor head.
We compute two permutation-invariant summaries of the per-atom features \(\{h_i(t)\}_{i=1}^N\):
a feature-wise attention pool \(p^{(1)}\) and a uniform mean \(p^{(2)}\).
Let \(a(h_i)\in\mathbb{R}^D\) be the output of a learned scoring MLP applied to \(h_i\).
For each feature channel \(d\),
\begin{equation}
    W_{i,d} \;=\; \frac{\exp(a_d(h_i))}{\sum_{k=1}^N \exp(a_d(h_k))},\qquad
p^{(1)}_d \;=\; \sum_{i=1}^N W_{i,d}\,h_{i,d},
\end{equation}
and
\begin{equation}
    p^{(2)} \;=\; \frac{1}{N}\sum_{i=1}^N h_i.
\end{equation}
Two MLP heads produce scalars \(v_1=v_{\phi_1}\!\left(p^{(1)}\right)\) and \(v_2=v_{\phi_2}\!\left(p^{(2)}\right)\).
The final value is a gated sum
\[
V_\phi(s_t) \;=\; \sigma(w_1)\,v_1 \;+\; \sigma(w_2)\,v_2, \ \ \ \ \ \ \phi = (\phi_1, \phi_2) , 
\]
where \(w_1,w_2\in\mathbb{R}\) are learnable scalars updated by backprop through the value loss and
\(\sigma(\cdot)\) is the logistic sigmoid.
This dual value network strategy is inspired by twin-critic architectures~\citep{hasselt2010double,van2016deep} in which two value estimators are trained in parallel. Their outputs are combined via a learnable weighted sum, providing lower bias and improved stability compared to a single critic. This is conceptually related to Double Q-learning and Clipped Double Q-learning, but differs in that we learn the aggregation weights rather than fixing them.

\subsection{GAE and bootstrapping}
We collect trajectories until a target number of complete episodes is reached.
Let $d_t\in\{0,1\}$ be the terminal flag ($d_t = 1$ if $s_{t+1}$ is terminal).
Define \textcolor{black}{temporal difference} (TD) residuals
\begin{equation}
\delta_t \;=\; r_t + \gamma(1-d_t)\,V_\phi(s_{t+1}) - V_\phi(s_t),
\end{equation}
and generalized advantages (GAE)~\citep{schulman2015high} with parameter $\lambda\in[0,1]$,
truncated at episode ends,
\begin{equation}
\hat A_t \;=\; \sum_{\ell\ge 0} (\gamma\lambda)^\ell
\Bigg(\prod_{m=0}^{\ell-1}(1-d_{t+m})\Bigg)\,\delta_{t+\ell},
\qquad
\hat R_t \;=\; \hat A_t + V_\phi(s_t).
\end{equation}
Equivalently, the backward recursion $\hat A_t=\delta_t+\gamma\lambda(1-d_t)\hat A_{t+1}$ with $\hat A_T=0$.
We normalize $\{\hat A_t\}$ per episode. \\
At each step $t$ we append a record $\tau_t=\big(A_t,\ i_t,\ j_t,\ r_t,\ d_t,\ t,\ H,\ \log\pi^{(a)}_{\text{old}}(i_t \mid s_t),\ \log\pi^{(p)}_{\text{old}}(j_t \mid s_t,i_t),\ V_{\phi,\text{old}}(s_t)\big)$,
where $S_t$ is the state before the swap, $(i_t,j_t)$ is the sampled (anchor, partner), $r_t$ the reward, $d_t \in \{0,1\}$ the terminal flag, $(t,H)$ reconstruct the time code, $\log\pi^{(a)}_{\text{old}},\log\pi^{(p)}_{\text{old}}$ are the detached headwise log-probs under the behavior policy, and $V_{\phi,\text{old}}(s_t)$ the detached critic value. We also cache the joint old log-prob $\log\pi_{\text{old}}=\log\pi^{(a)}_{\text{old}}+\log\pi^{(p)}_{\text{old}}$. After collecting a target number of complete episodes, we compute $(\hat A_t,\hat R_t)$ and, during PPO updates, re-encode each stored $\hat A_t$ with the current networks to obtain current logits/entropies, rebuild the same mask from atomic numbers, reconstruct $\psi(t,H)$ from $(t,H)$, and form the current joint log-prob $\log\pi_\theta=\log\pi^{(a)}_\theta+\log\pi^{(p)}_\theta$ used in the importance ratio and KL gate.

\subsection{PPO objectives and loss}
Given the factorized policy of Equation~\eqref{eq:factored_policy}, the joint log-probability used for training is
\begin{equation}
\log \pi_\theta(a_t\mid s_t)
=
\underbrace{\log\mathrm{softmax} \big(
\pi_{\theta_a}\left(i \mid s_t\right)
\big)}_{\text{anchor}}
+
\underbrace{\log\mathrm{softmax} \big(
\pi_{\theta_p}\left(j \mid s_t, i\right)
\big)}_{\text{partner}}.
\label{eq:jointlog}
\end{equation}
Using the joint log-probability is necessary because the sampled action $a_t=(i_t,j_t)$ comes from a product distribution with a conditional partner head; equivalently,
\begin{equation}
    \frac{\pi_\theta(a_t\mid s_t)}{\pi_{\theta_{\text{old}}}(a_t\mid s_t)}
=
\frac{\pi_{\theta_a}(i_t\mid s_t)}{\pi_{\theta_a^{\text{old}}}(i_t\mid s_t)}
\cdot
\frac{\pi_{\theta_p}(j_t\mid s_t,i_t)}{\pi_{\theta_p^{\text{old}}}(j_t\mid s_t,i_t)}.
\end{equation}
With this, we define the importance ratio
\begin{equation}
\rho_t(\theta) \;=\; \frac{\pi_\theta(a_t\mid s_t)}{\pi_{\theta_{\text{old}}}(a_t\mid s_t)}
= \exp \Big(\log\pi_\theta(a_t\mid s_t)-\log\pi_{\theta_{\text{old}}}(a_t\mid s_t)\Big).
\label{eq:imp_ratio}
\end{equation}
Following PPO~\citep{schulman2017proximal}, we optimize the clipped surrogate
\begin{equation}
\mathcal{L}^{\mathrm{CLIP}}(\theta)=
\mathbb{E}_t\Big[\min\big(\rho_t(\theta)\,\hat A_t,\ \mathrm{clip}(\rho_t(\theta),\,1-\epsilon,\,1+\epsilon)\,\hat A_t\big)\Big],
\label{eq:clip}
\end{equation}
where $\epsilon>0$ is the trust-region width and $\hat A_t$ is the (per-episode normalized) generalized advantage estimate~\citep{schulman2015high}. Here we have taken an $\epsilon$-greedy strategy ($\epsilon=0.05$).The clipping term is an unconstrained proxy for the trust-region penalty~\citep{schulman2015trust}, limiting harmful updates when $\rho_t$ deviates too far from $1$ with $\hat A_t\neq 0$. We then combine the policy surrogate with a value regression and entropy regularization and minimize
\begin{equation}
\mathcal{J}(\theta,\phi)=
-\alpha_{\text{pol}}\mathcal{L}^{\mathrm{CLIP}}(\theta)
+\alpha_{\text{vf}}\,\mathbb{E}_t\big[\ell_{\text{SmoothL1}}(V_\phi(s_t),\hat R_t)\big]
-\alpha_{\text{ent}}\,\mathbb{E}_t\big[H(\pi_{\theta_a}(\cdot\mid s_t))+H(\pi_{\theta_p}(\cdot\mid s_t,i_t))\big],
\label{eq:fullobj}
\end{equation}
with coefficients $\alpha_{\text{pol}},\alpha_{\text{vf}},\alpha_{\text{ent}}>0$. Here $\hat R_t=\hat A_t+V_\phi(s_t)$ is the bootstrap target and $\ell_{\text{SmoothL1}}$ is the Huber loss~\citep{huber1992robust}, chosen for robustness to occasional large post-relaxation energy drops (heavy-tailed targets). Entropies $H(\cdot)$ are computed on the masked categoricals and encourage exploration in both heads~\citep{williams1992simple,mnih2016asynchronous}.

\paragraph{KL gating (early stopping).}
To further bound policy drift across PPO epochs, we monitor the empirical \emph{joint} KL on each mini-batch,
\begin{equation}
\widehat{\mathrm{KL}}(\pi_{\text{old}}\Vert \pi_\theta)
=\mathbb{E}_t\big[\log\pi_{\text{old}}(a_t\mid s_t)-\log\pi_\theta(a_t\mid s_t)\big],
\label{eq:kl}
\end{equation}
computed with Equation~\eqref{eq:jointlog} over the masked feasible set. If $\widehat{\mathrm{KL}} > c_{\text{stop}}\cdot \tau$ with $c_{\text{stop}}=1.5$ and $\tau$ being a target max KL divergence (a hyperparameter set to 0.015 in our experiments), we stop the current PPO epoch early (break the mini-batch loop). This empirically prevents collapse and aligns with the trust-region motivation of TRPO/PPO~\citep{schulman2015trust,schulman2017proximal}.

\clearpage
\section{Training curves} \label{app:training_curves}
\begin{figure}[ht!]
  \centering
  \begin{subfigure}[t]{0.49\textwidth}
    \centering\includegraphics[width=\textwidth]{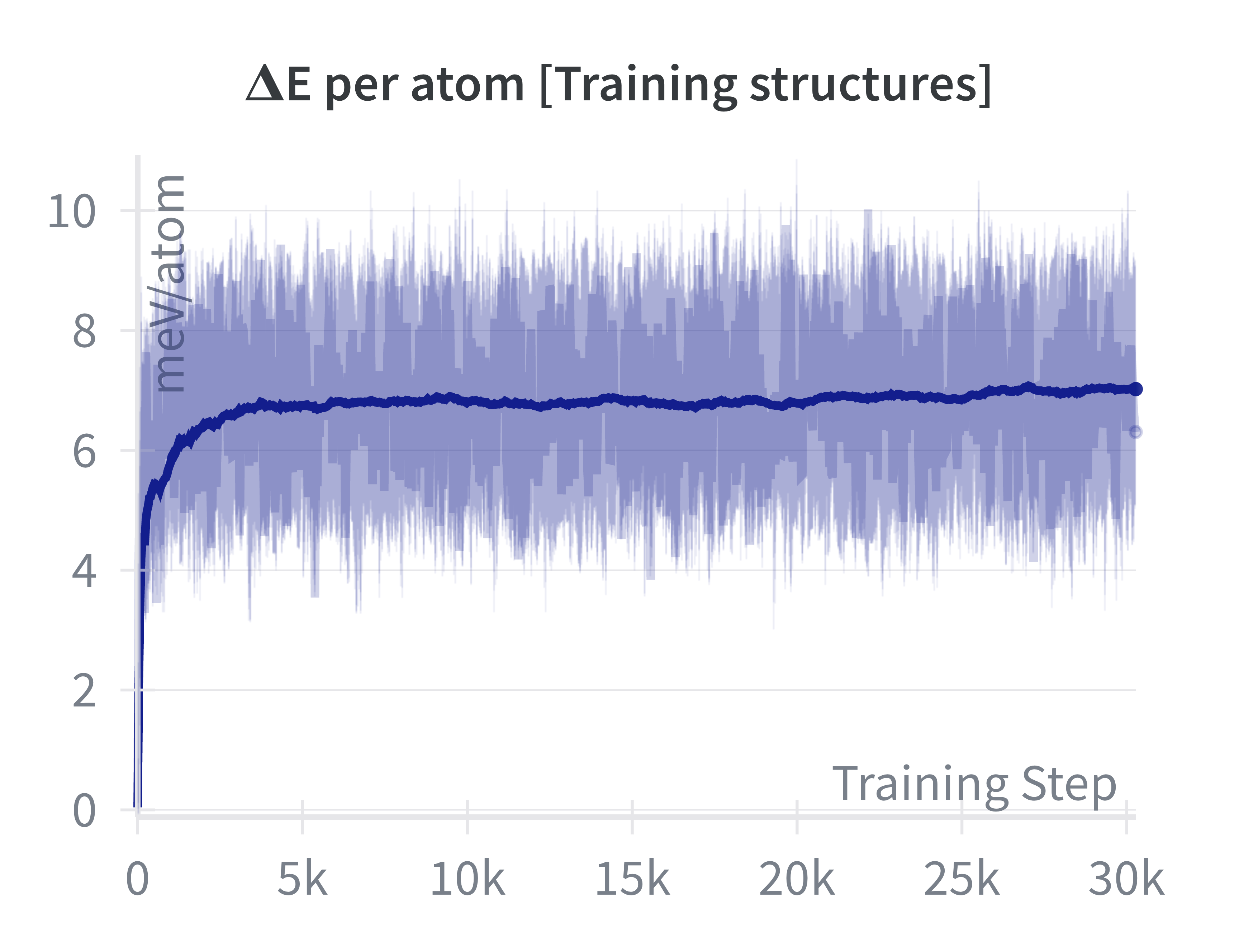}
    \caption{Training progress for the energy reduction per atom over the full training horizon, in $\mathrm{meV/atom}$}\label{fig:deltae_training}
  \end{subfigure}\hfill
  \begin{subfigure}[t]{0.49\textwidth}
    \centering\includegraphics[width=\textwidth]{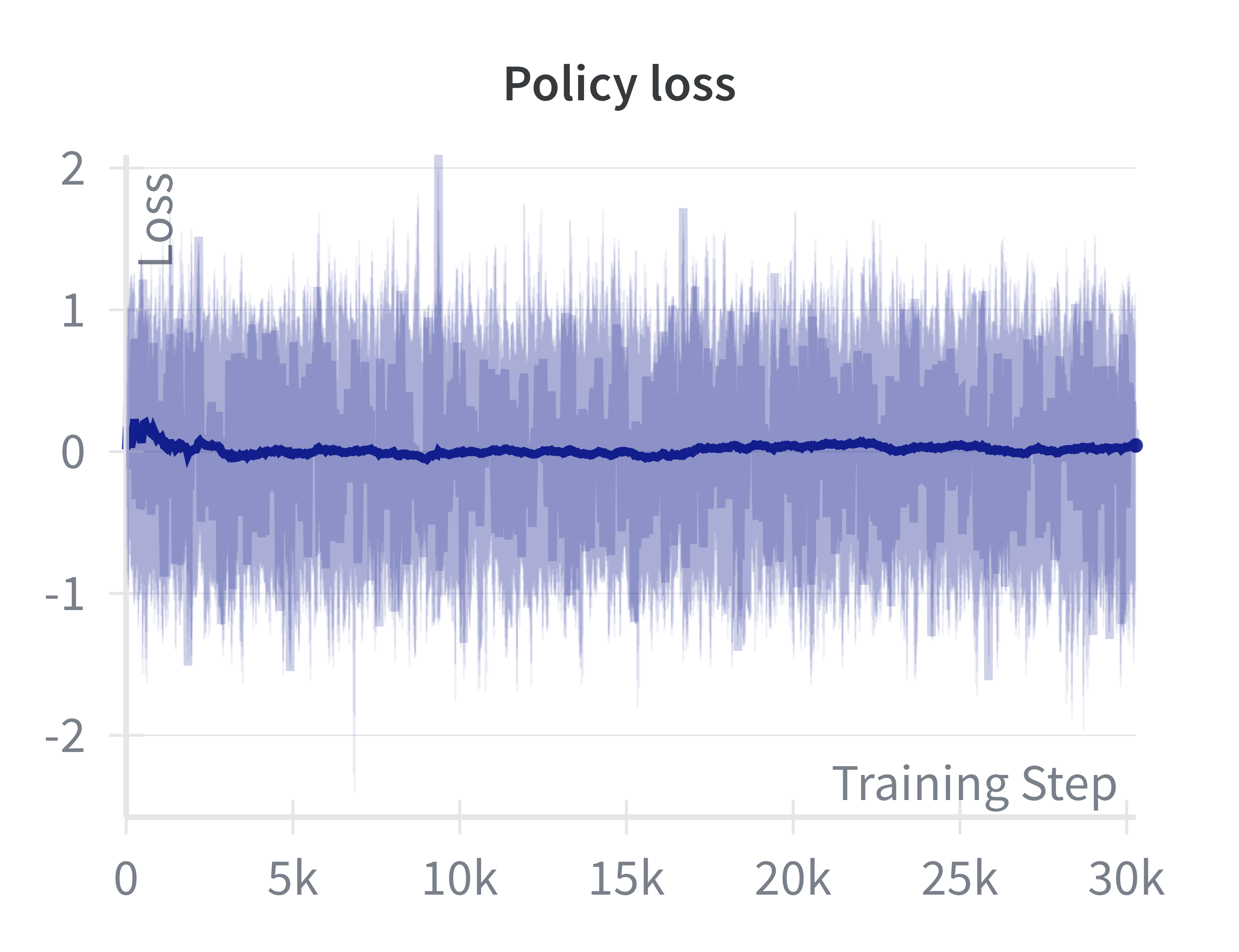}
    \caption{Training progress for the policy loss}\label{fig:pol_loss}
  \end{subfigure}
  \medskip
  \medskip
  \medskip
  \medskip
  \begin{subfigure}[t]{0.49\textwidth}
    \centering\includegraphics[width=\textwidth]{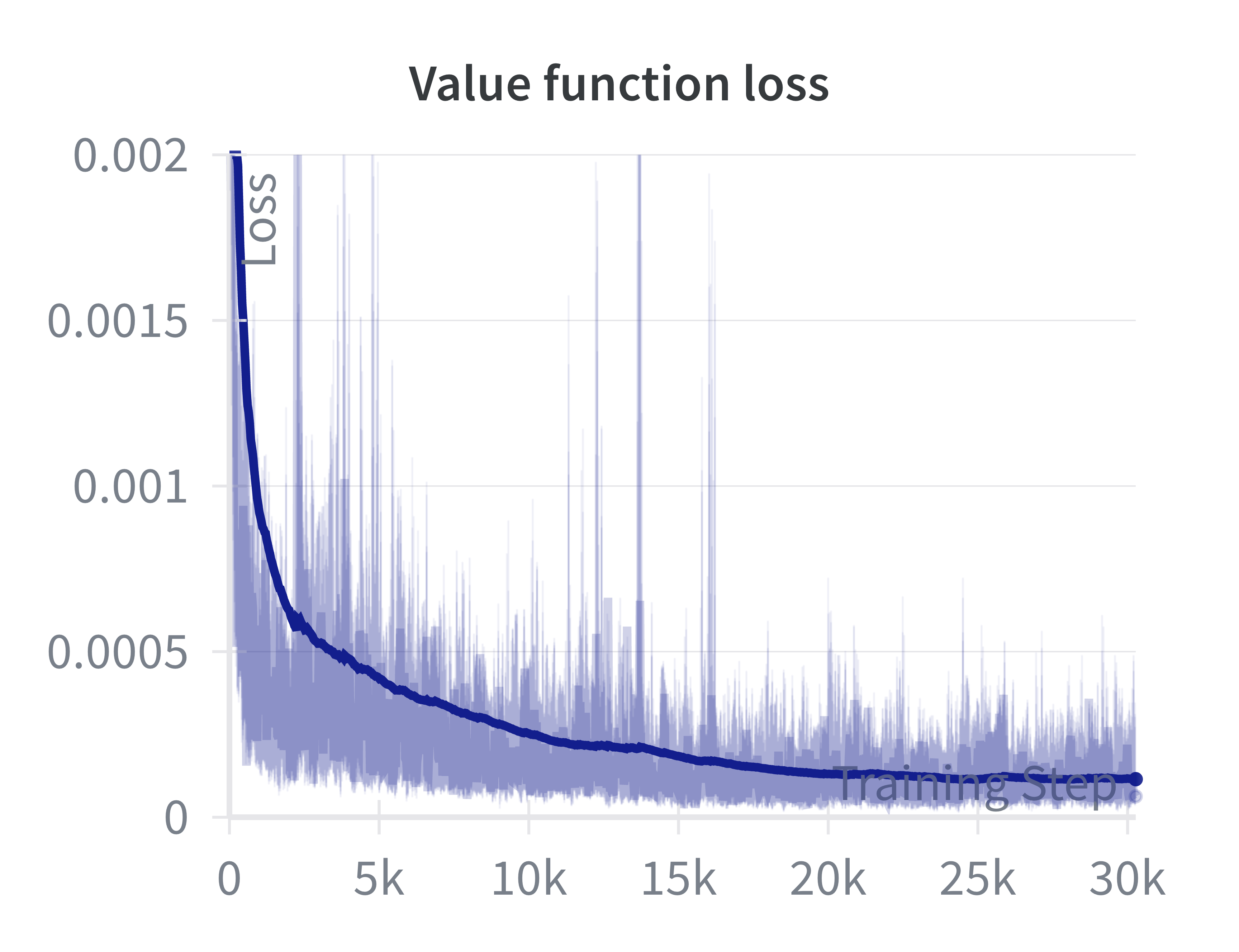}
    \caption{Training progress for the value function loss}\label{fig:val_loss}
  \end{subfigure}\hfill
  \begin{subfigure}[t]{0.49\textwidth}
    \centering\includegraphics[width=\textwidth]{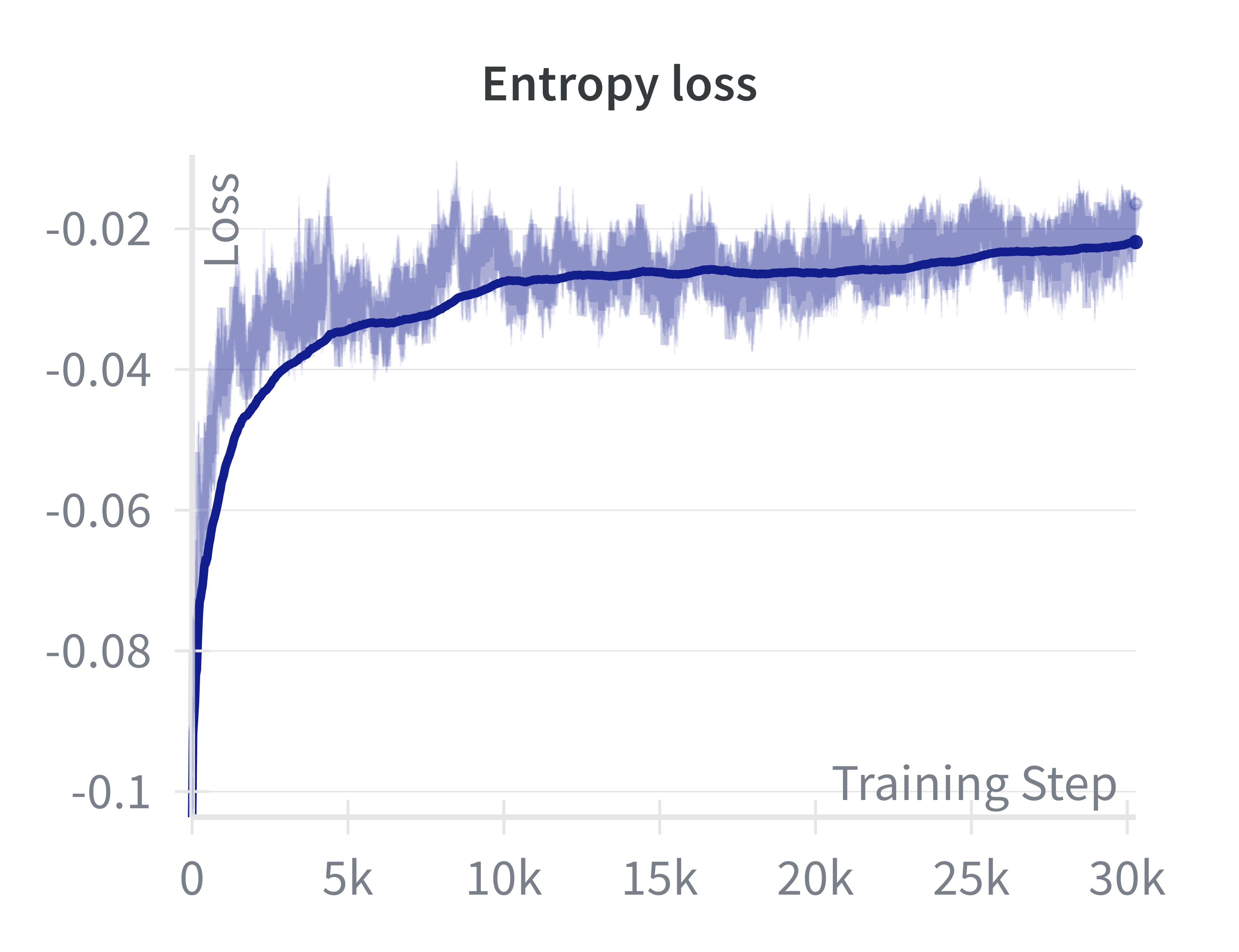}
    \caption{Training progress for the entropy loss}\label{fig:ent_loss}
  \end{subfigure}
  \caption{Training curves for our RL agent \textcolor{black}{for Experiment 1}. \textbf{(a)}: The pattern is typical for a well-posed PPO procedure, with return rising over time. In this case, we display the final energy reduction per atom, which is proportional to the return $G$, i.e. $\frac{\Delta E}{atom}=\frac{E_0-E_H}{N_{atoms}}=\frac{G}{N_{atoms}}$ (see Equation \ref{eq:telescoping}). \textbf{(b)}: Since we use advantage normalization (GAE), the expectation value of the advantage $\hat A$ is $\mathbb{E}[ \hat A] \approx 0$, and near a local optimum the importance ratio (Equation \ref{eq:imp_ratio}) is $\rho_{t} \approx 1$, so the expected surrogate is near zero. Furthermore, clipping plus KL gating keeps $\rho_t$ close to 1, which further dampens any trend. Finite-sample noise, changing data (on-policy), and alternating updates of policy/critic then produce batches where $\mathcal{L}^{\text {CLIP }}$ is slightly positive or negative. Thus, the observed oscillation around 0 is expected. \textbf{(c)}: Meanwhile, the value loss falls because $V_\phi$ learns to match the bootstrap targets, shrinking TD errors over time. \textbf{(d)}: As the policy sharpens, the entropy $H(\pi)$ decreases as the policy \textcolor{black}{becomes} more exploitative. }
  \label{fig:training_curves}
\end{figure}

\begin{figure}[ht!]
  \centering
  \begin{subfigure}[t]{0.49\textwidth}
    \centering\includegraphics[width=\textwidth]{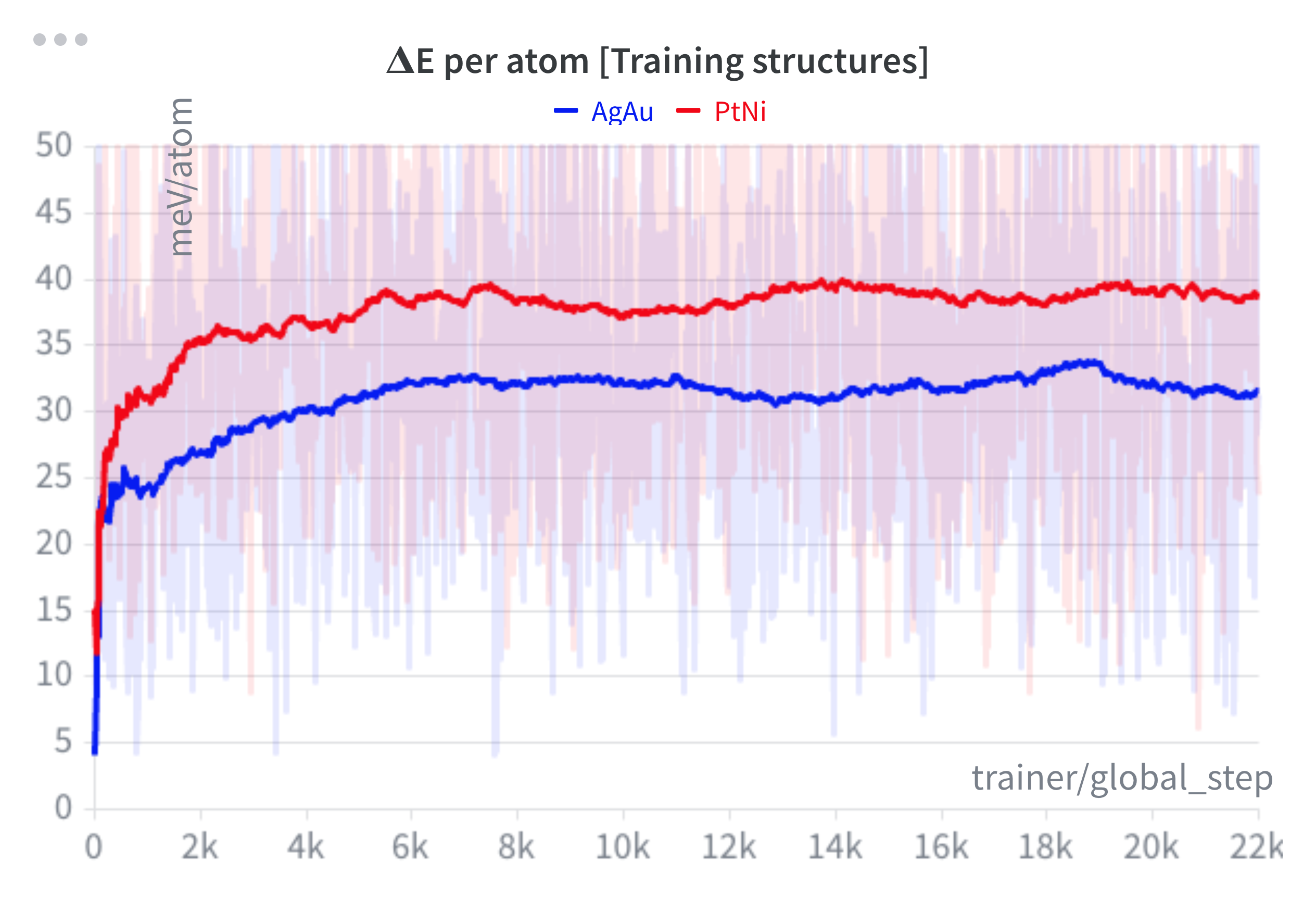}
    \caption{Training progress for the energy reduction per atom over the full training horizon, in $\mathrm{meV/atom}$}\label{fig:deltae_training_exp3}
  \end{subfigure}\hfill
  \begin{subfigure}[t]{0.49\textwidth}
    \centering\includegraphics[width=\textwidth]{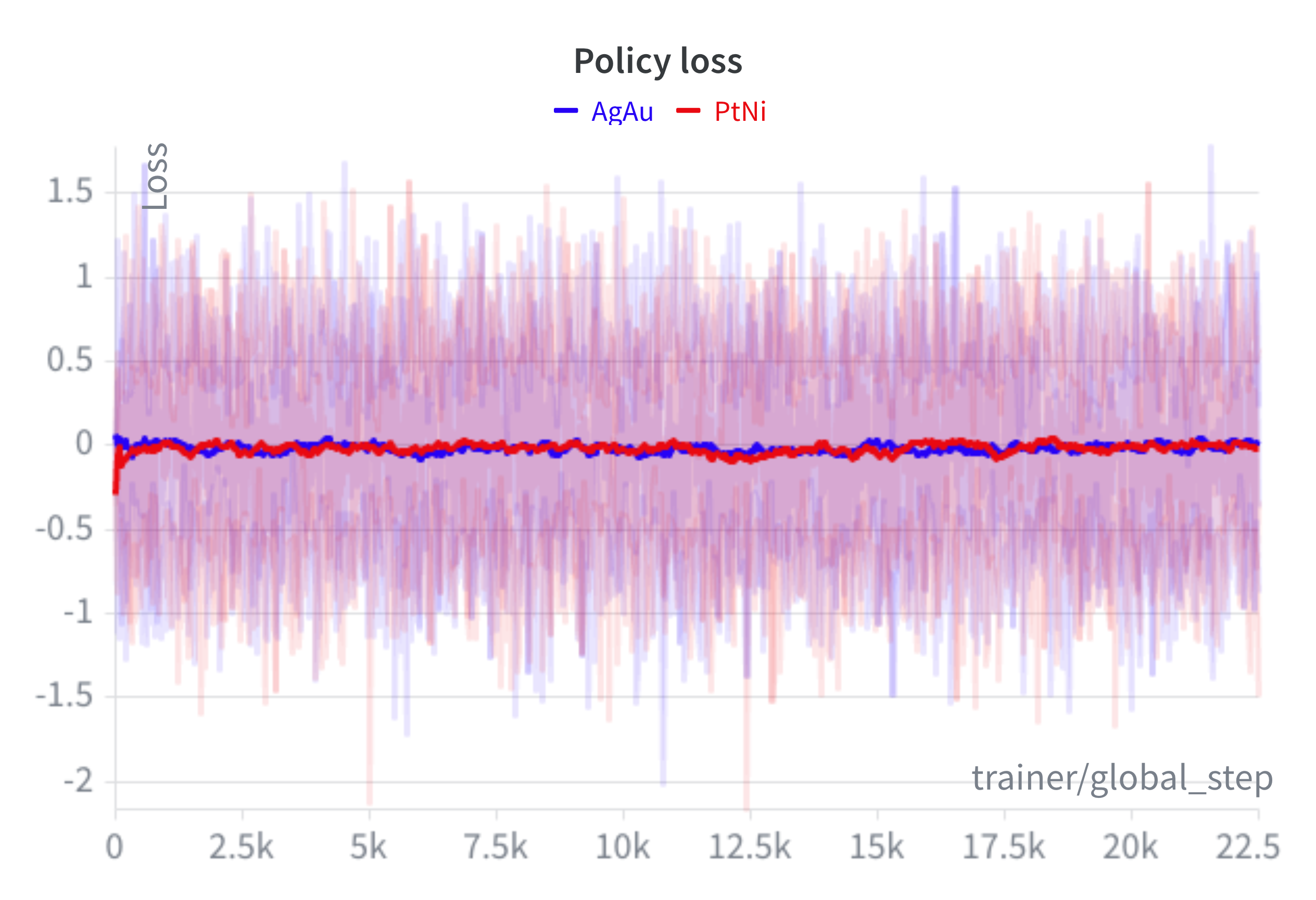}
    \caption{Training progress for the policy loss}\label{fig:pol_loss_exp3}
  \end{subfigure}
  \medskip
  \medskip
  \medskip
  \medskip
  \begin{subfigure}[t]{0.49\textwidth}
    \centering\includegraphics[width=\textwidth]{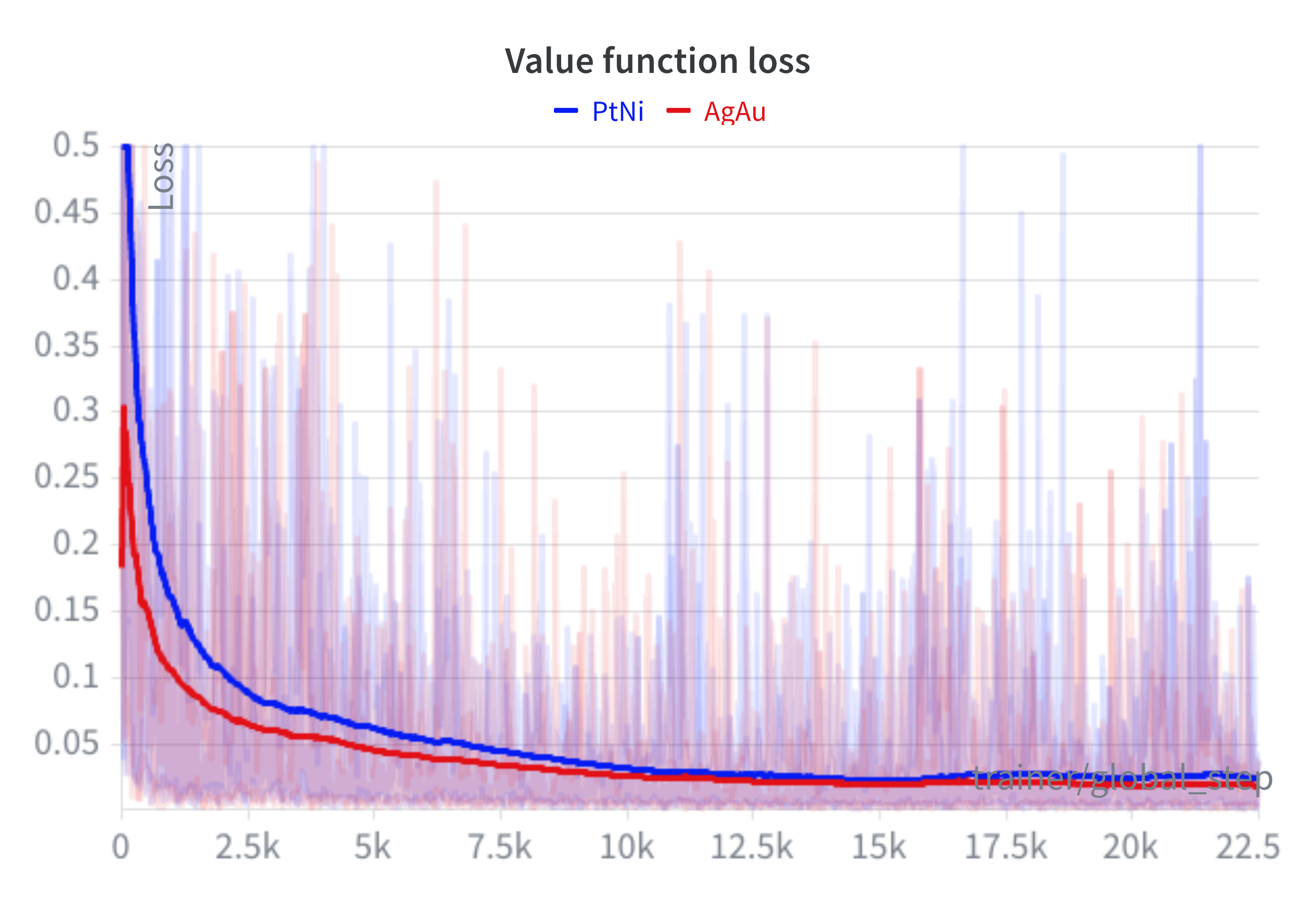}
    \caption{Training progress for the value function loss}\label{fig:val_loss_exp3}
  \end{subfigure}\hfill
  \begin{subfigure}[t]{0.49\textwidth}
    \centering\includegraphics[width=\textwidth]{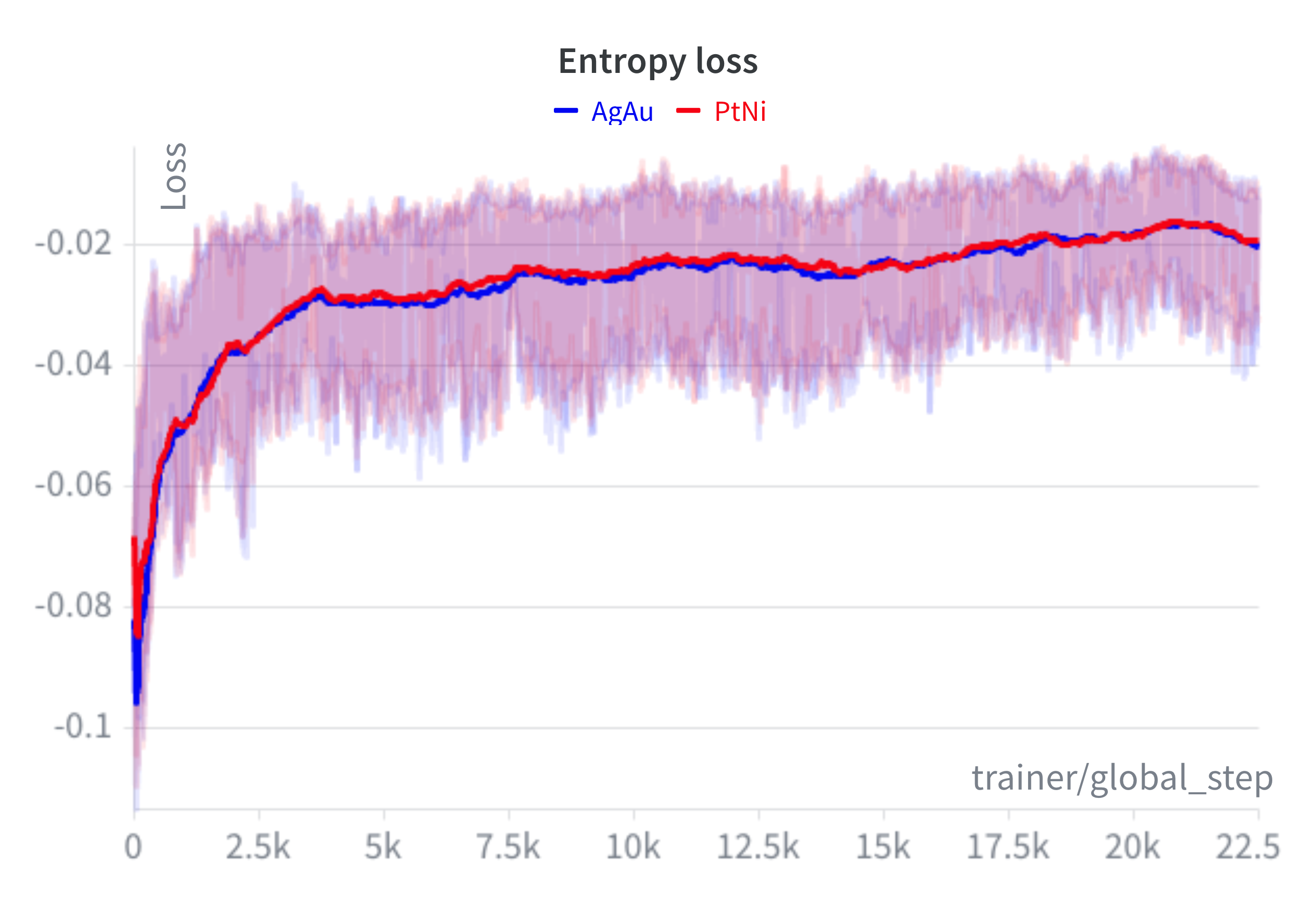}
    \caption{Training progress for the entropy loss}\label{fig:ent_loss_exp3}
  \end{subfigure}
  \caption{\textcolor{black}{Training curves for the RL agent for Experiment 3. The agent samples both AgAu and PtNi compositions, but we show the training here with separate curves for each type of system.}}
  \label{fig:training_curves_exp3}
\end{figure}
\clearpage
\section{Additional results} \label{app:additional results}
\subsection{Full trajectory example}
\begin{figure}[ht!]
  \centering
  % Row 1
  \begin{subfigure}[t]{0.25\textwidth}
    \centering\includegraphics[width=\textwidth]{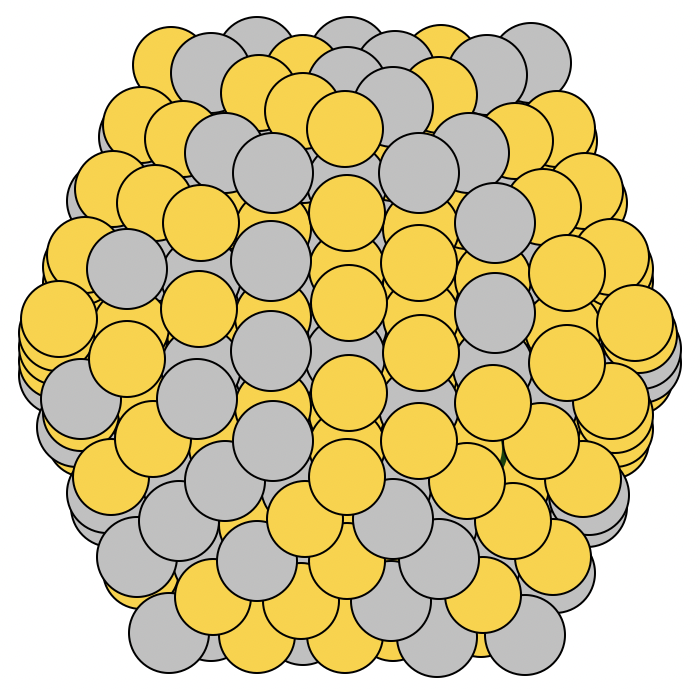}
    \caption*{$t=0$}\label{fig:t0}
  \end{subfigure}\hfill
  \begin{subfigure}[t]{0.25\textwidth}
    \centering\includegraphics[width=\textwidth]{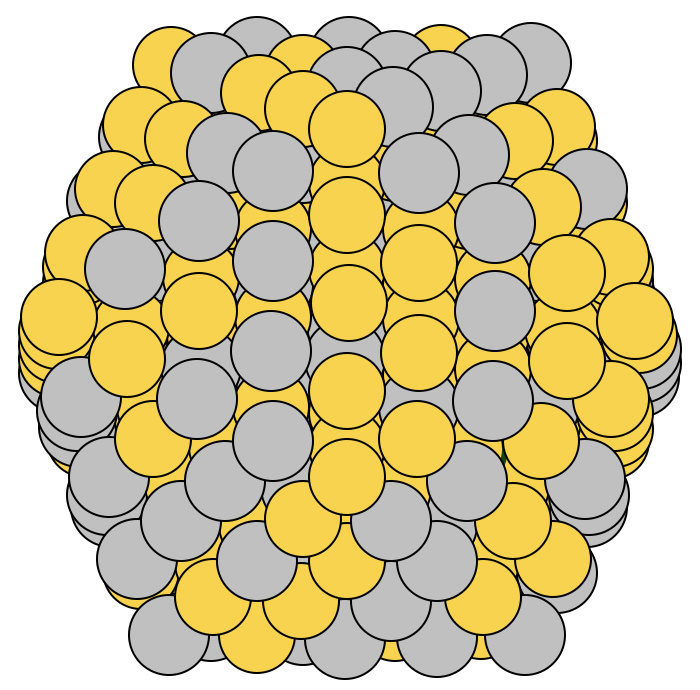}
    \caption*{$t=15$}\label{fig:t15}
  \end{subfigure}\hfill
  \begin{subfigure}[t]{0.25\textwidth}
    \centering\includegraphics[width=\textwidth]{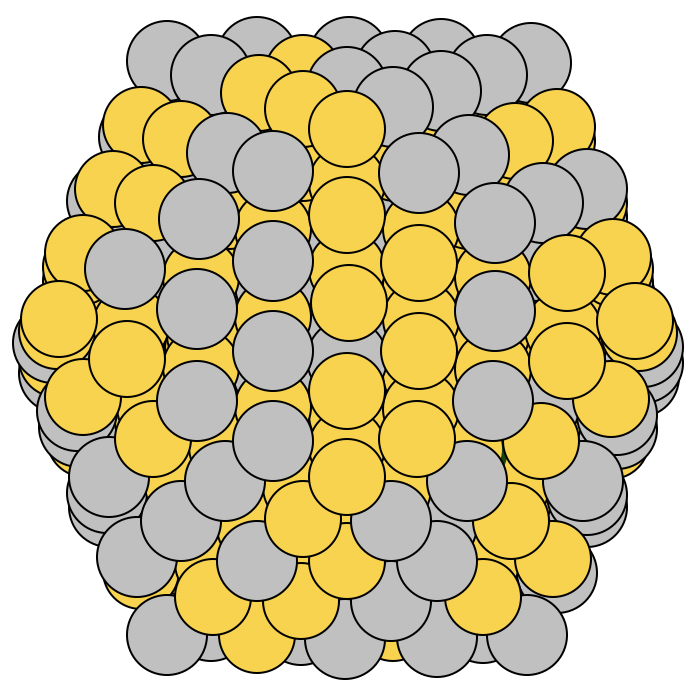}
    \caption*{$t=30$}\label{fig:t30}
  \end{subfigure}

  \medskip
  % Row 2
  \begin{subfigure}[t]{0.25\textwidth}
    \centering\includegraphics[width=\textwidth]{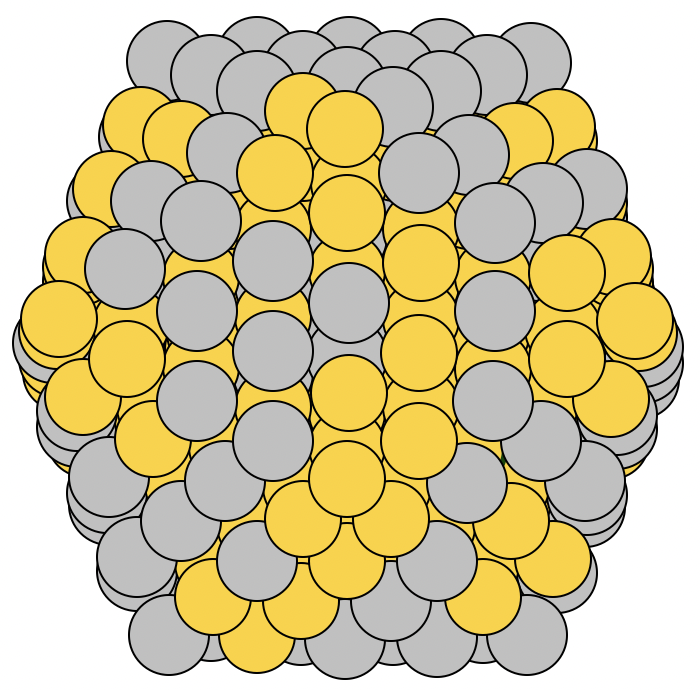}
    \caption*{$t=45$}\label{fig:t45}
  \end{subfigure}\hfill
  \begin{subfigure}[t]{0.25\textwidth}
    \centering\includegraphics[width=\textwidth]{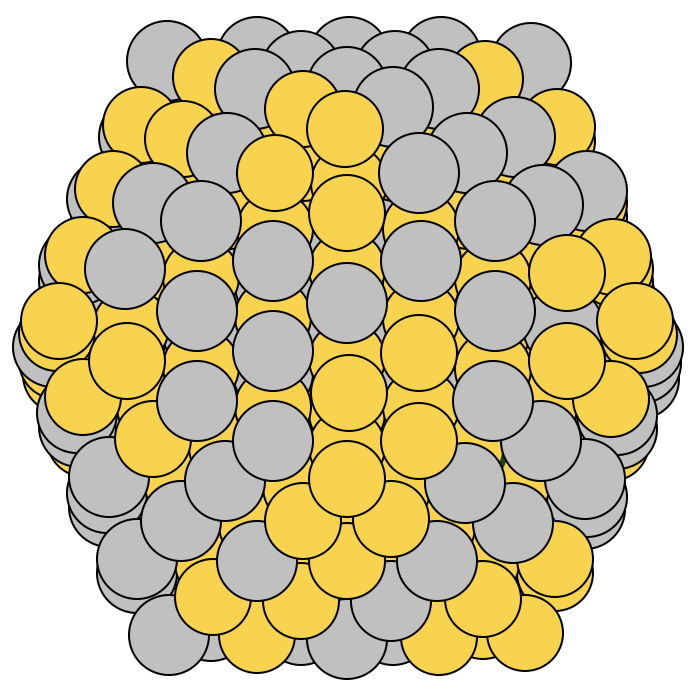}
    \caption*{$t=60$}\label{fig:t60}
  \end{subfigure}\hfill
  \begin{subfigure}[t]{0.25\textwidth}
    \centering\includegraphics[width=\textwidth]{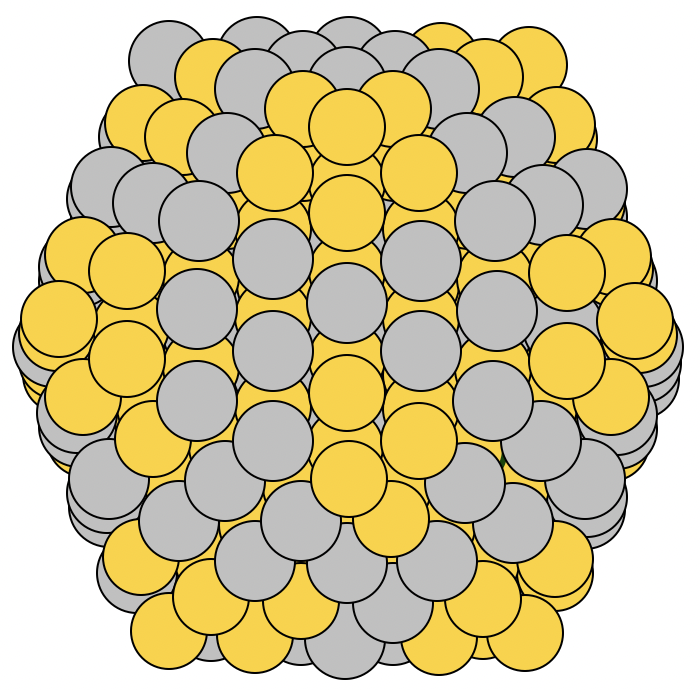}
    \caption*{$t=75$}\label{fig:t75}
  \end{subfigure}

  \medskip
  % Row 3
  \begin{subfigure}[t]{0.25\textwidth}
    \centering\includegraphics[width=\textwidth]{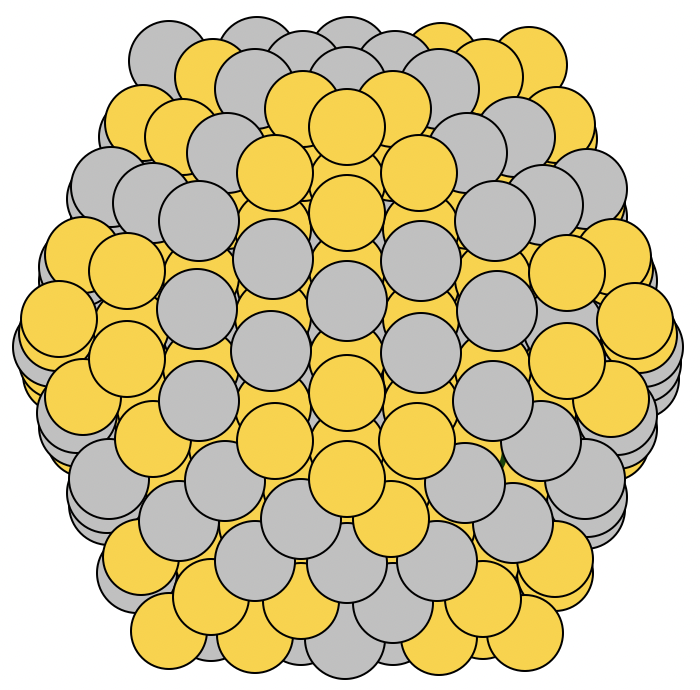}
    \caption*{$t=90$}\label{fig:t90}
  \end{subfigure}\hfill
  \begin{subfigure}[t]{0.25\textwidth}
    \centering\includegraphics[width=\textwidth]{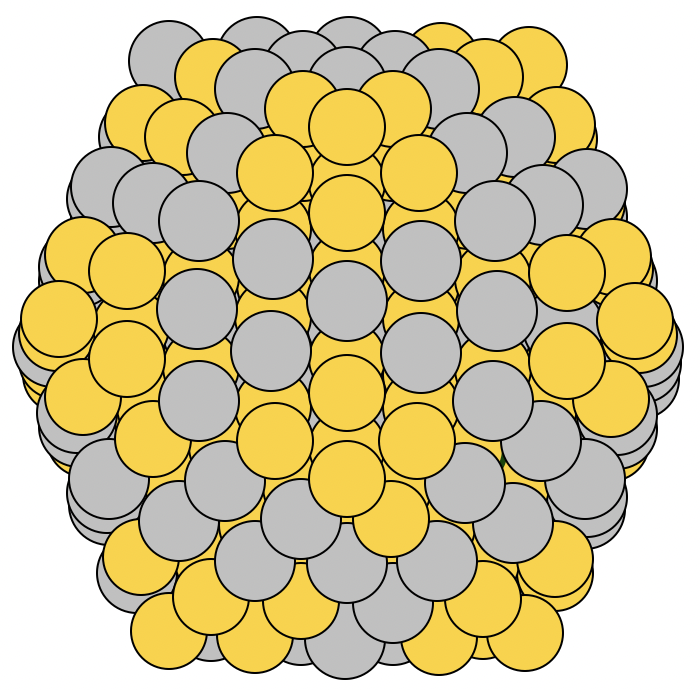}
    \caption*{$t=105$}\label{fig:t105}
  \end{subfigure}\hfill
  \begin{subfigure}[t]{0.25\textwidth}
    \centering\includegraphics[width=\textwidth]{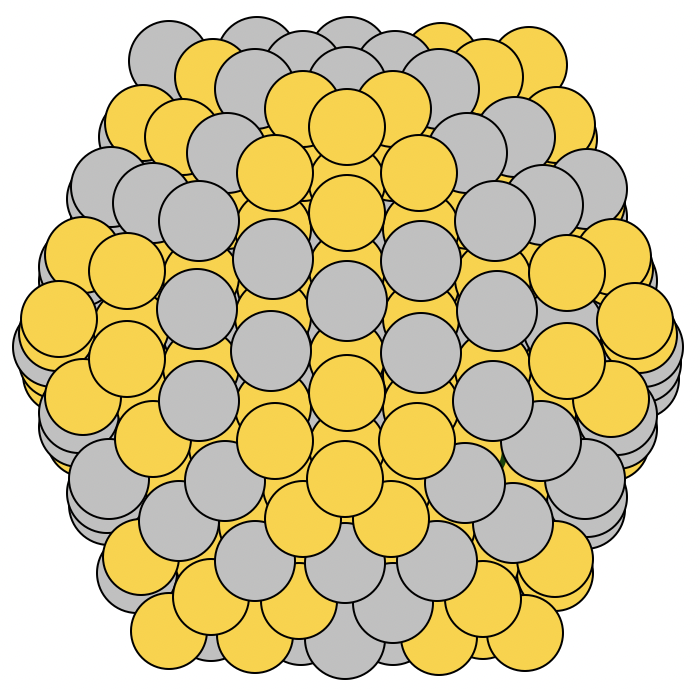}
    \caption*{$t=120$}\label{fig:t120}
  \end{subfigure}

  \medskip
  % Row 4
  \begin{subfigure}[t]{0.25\textwidth}
    \centering\includegraphics[width=\textwidth]{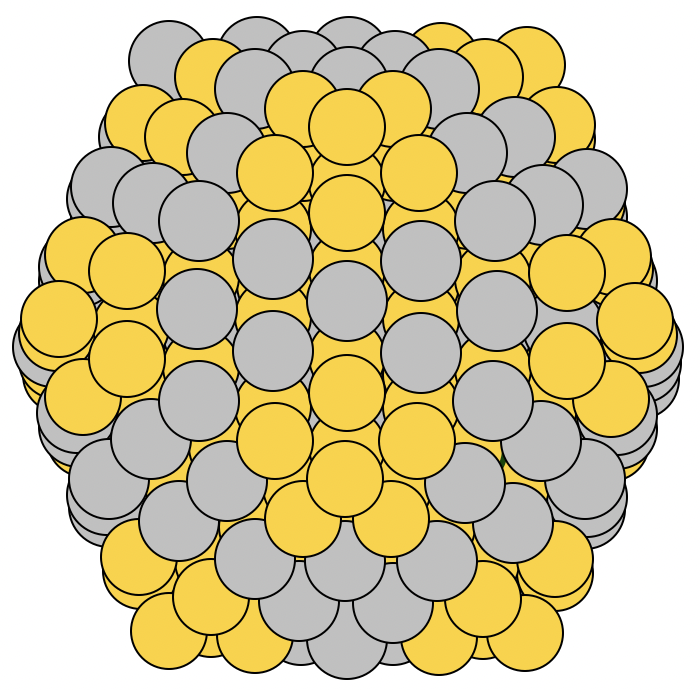}
    \caption*{$t=135$}\label{fig:t135}
  \end{subfigure}\hfill
  \begin{subfigure}[t]{0.25\textwidth}
    \centering\includegraphics[width=\textwidth]{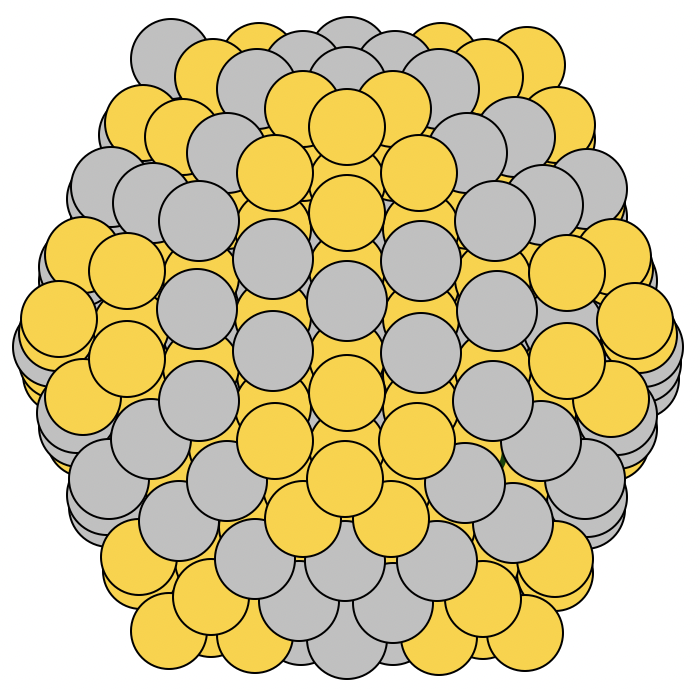}
    \caption*{$t=150$}\label{fig:t150}
  \end{subfigure}\hfill
  \begin{subfigure}[t]{0.25\textwidth}
    \centering\includegraphics[width=\textwidth]{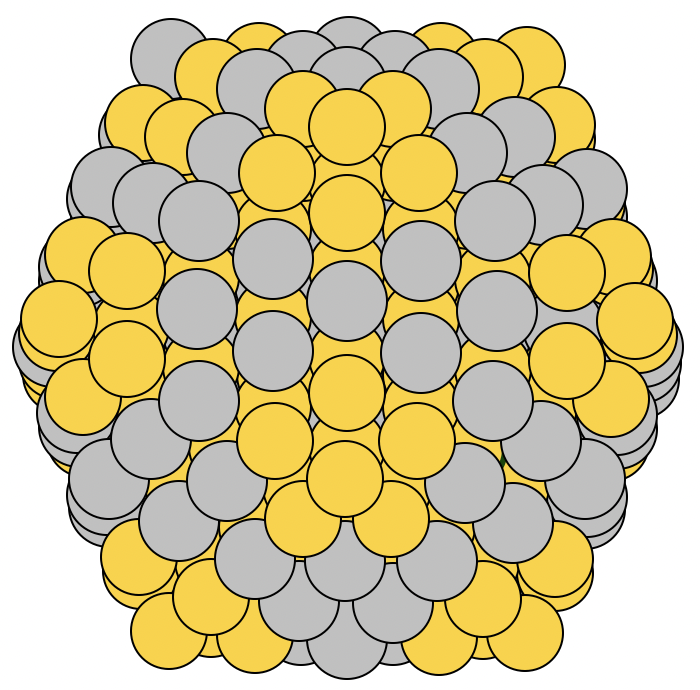}
    \caption*{$t=165$}\label{fig:t165}
  \end{subfigure}

  \caption{Snapshots from the first 165 steps (atomic swaps) for the best \ch{Ag126Au183} configuration found by the RL agent, which is equivalent to the provably optimal ordering found in~\citep{larsen2018rich}. As seen in the trajectory, most of the atoms in the final solution of Figure~\ref{subfig:ih5} are already in their final position at $t=165$, even though the evaluation horizon is $H=2\times N_{atoms}=618$. In particular, the top-left \ch{Ag} atom is still out of place at $t=165$ compared to Figure \ref{subfig:ih5}, but the agent swaps this atom for \ch{Au} around $t=300$. \textcolor{black}{All structures depicted in this figure are made with ASE~\citep{larsen2017atomic}.}}
  \label{fig:full_traj_snaps}
\end{figure}
\clearpage
\subsection{\textcolor{black}{Experiment 3 PtNi final structures}}
\begin{figure}[ht]
    \centering
    % Row 1
    \begin{subfigure}{0.49\textwidth} % top-left cell
        \centering
        \includegraphics[width=0.7\linewidth]{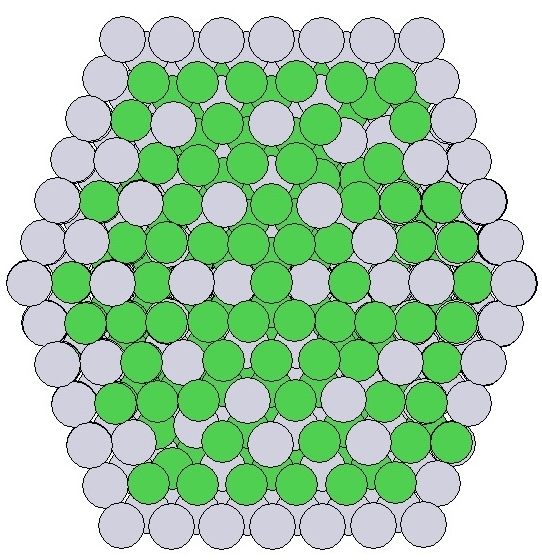}
        \caption*{\ch{Ni113Pt196}}
        \label{subfig:ptnia}
    \end{subfigure}
    \begin{subfigure}{0.49\textwidth} % top-right cell
        \centering
        \includegraphics[width=0.7\linewidth]{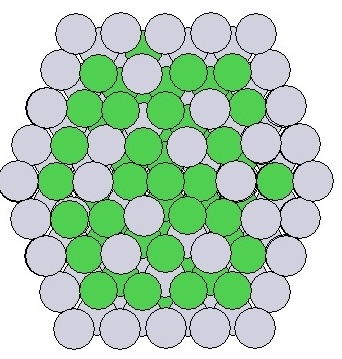}
        \caption*{\ch{Ni419Pt504}}
        \label{subfig:ptnib}
    \end{subfigure}    
    % Row 2
    \caption{\textcolor{black}{In Experiment 3, PtNi nanoparticle chemical orderings predicted by the jointly trained policy (AgAu and PtNi) are similar to the structures reported in the literature for \ch{Ni113Pt196} and \ch{Ni419Pt504} icosahedral nanoparticles. The RL model correctly recovers the Pt-rich outer shell in both cases, supporting that the jointly trained policy captures the physically expected surface segregation trend in PtNi nanoalloys, although there are discrepancies in the inner shell elemental ordering. Both structures depicted in this figure are made with ASE~\citep{larsen2017atomic}.}}
    \label{fig:PtNi}
\end{figure}
%\subsection{Initialization robustness} \label{app:init_dependency}
% in preamble:
% \usepackage{graphicx}
% \usepackage{subcaption}
% \captionsetup[sub]{labelfont=bf,textfont=normalfont}
% in preamble:
% \usepackage{graphicx}

% In your preamble:
% \usepackage{graphicx}
% \usepackage{subcaption}

\newpage
\bibliographystyle{unsrtnat} %NEW
\bibliography{references}    % references.bib must exist

%\newpage
%\begin{figure}[t]
  %  \centering
    %\includegraphics[width=8.10cm]{figs/TOC_graphic/TOC %Graphic.png}
   % \caption{TOC Graphic}
%\end{figure}

\end{document}